\documentclass[%
aip,
author-year,%
author-numerical,%
]{revtex4-2}
\usepackage{amsmath,latexsym,amssymb}
\usepackage[%
  colorlinks=true,
  urlcolor=red,
  linkcolor=blue,
  citecolor=green
]{hyperref}
\usepackage{etoolbox}
\usepackage{breqn}

\makeatletter
\let\cat@comma@active\@empty
\makeatother

\usepackage{graphicx}
\usepackage{dcolumn}
\usepackage{bm}
\usepackage{float}
\usepackage{graphicx}
\usepackage{subfig}
\usepackage{ntheorem}

\theoremstyle{nonumberplain}

\usepackage{hyperref}
\hypersetup{colorlinks,
citecolor=blue
}

\begin{document}
\title{Gravitational lensing in Brill spacetimes}

\author{Mourad Halla}
\email{mourad.halla@zarm.uni-bremen.de.}
\author{Volker Perlick}%
 \email{perlick@zarm.uni-bremen.de.}
\affiliation{ZARM, University of Bremen, 28359 Bremen, Germany}%
\date{\today}

\begin{abstract}
We consider the Brill metric which is an electrovacuum solution to
Einstein's field equation. It depends on three parameters, a mass
parameter $m$, a NUT parameter $l$ and a charge parameter $e$. If
the charge parameter is small, the metric describes a black hole;
if it is sufficiently big, it describes a wormhole. We determine the 
relevant lensing features both in the black-hole and in the wormhole 
case. In particular, we give formulas for the photon spheres, for the 
angular radius of the shadow and for the deflection angle. We illustrate 
the lensing features with the help of an effective potential and in 
terms of embedding diagrams. To that end we make use of the fact that 
each lightlike geodesic is contained in a (coordinate) cone and that 
it is a geodesic of a Riemannian optical metric on this cone. By the
Gauss-Bonnet theorem, the sign of the Gaussian curvature of the 
optical metric determines the sign of the deflection angle. In the 
wormhole case the deflection angle may be negative which means that 
light rays are repelled from the center.
\end{abstract}
\maketitle

\section{INTRODUCTION}
Gravitational lensing is one of the most important tools for observing 
(ultra-)compact objects such as black holes or wormholes. In this case
the weak-field small-angle approximation that is often employed in
lensing is not applicable because light rays can make arbitrarily 
many turns around the central object. Then one has to use the full
spacetime formalism of general relativity, without approximation, for
determining the lensing features, see e.g. the living review by 
Perlick \cite{Perlick2004}.

In this paper we want to apply this formalism to lensing in the Brill
spacetime which is an exact solution to the Einstein-Maxwell equations,
found by Brill \cite{Brill} in 1964. The Brill metric depends on a mass 
parameter $m$, a NUT parameter $l$ and a charge parameter $e$. For $l = 0$
the metric reduces to the Reissner-Nordstr{\"o}m metric which is static 
and spherically symmetric. As the light rays in the 
Reissner-Nordstr{\"o}m metric have been extensively discussed (see e.g. Chadrasekhar
\cite{Chandrasekhar1983}), we will restrict our investigation to the case 
$l \neq 0$. Then the metric is still stationary and it still admits an 
$SO (3, \mathbb{R})$ symmetry; however, it is no longer static and it is 
not spherically symmetric in the usual sense because the orbits of the 
$SO(3,\mathbb{R})$ symmetry are 3-dimensional timelike hypersurfaces rather 
than 2-dimensional spacelike spheres. For $e=0$ the Brill metric reduces 
to the Newman-Unti-Tamburino (NUT) vacuum solution \cite{NewmanTamburinoUnti1963} which describes 
a black hole. For non-zero $e$, the Brill metric still describes a black 
hole (now with charge) as long as $e^2$ is small, but for sufficiently 
big $e^2$ it describes a traversable wormhole, see Cl{\'e}ment et 
al. \cite{Clement}. These Brill wormholes are the only known traversable 
wormhole solutions to Einstein's field equation (in 4 spacetime dimensions) 
with an energy-momentum tensor that satisfies all energy conditions. We 
believe that for this reason it is worthwhile to study their lensing 
features in detail. 

The paper is organized as follows. In Sec. \ref{sec:Brill} we review the
basic features of Brill spacetimes. In Sec. \ref{sec:geo} and Sec. \ref{sec:light}
we derive, respectively, the relevant equations for general geodesics and for 
lightlike geodesics. We discuss the lensing features of 
Brill black holes in Sec. \ref{black} and of Brill wormholes in 
Sec. \ref{NUTW}. 

\section{Brill spacetimes}\label{sec:Brill}
The Brill metric, also known as the Reissner-Nordstr{\"o}m-NUT metric, is an exact solution of the Einstein-Maxwell  equations that was found by Brill in 1964 \cite{Brill} as a  generalization of the NUT metric \cite{NewmanTamburinoUnti1963}. It depends on three parameters, $m$, $l$ and $e$. In Boyer-Lindquist-type coordinates $(t, r, \vartheta, \varphi)$ the Brill metric reads   
\begin{equation}
\begin{aligned}
g_{\mu \nu} dx^{\mu} dx ^{\nu} &=-\dfrac{(r-m)^2+b}{r^2+l^2}\Big(dt-2l (\cos\vartheta+C)d\varphi \Big)^2\\
&+\frac{(r^2+l^2)dr^2}{(r-m)^2+b}+ (r^2+l^2)\Big(d\vartheta^2+\sin^2\vartheta\; d\varphi^2 \Big) \ ,
\end{aligned} 
\label{metric}
\end{equation}
with
\begin{equation}
b:= e^2-m^2-l^2 \ .
\end{equation}
$\vartheta$ and $\varphi$ are the standard coordinates on the two-sphere ${\mathbb{S}}^2$, whereas the time coordinate $t$ and the radial coordinate $r$ range over all of $\mathbb{R}$, unless in the case $l=0$ where the radial coordinate has to be restricted to $r>0$, or $r<0$, because there is a curvature singularity at $r=0$.

The metric is stationary, but not static, on the domain where $g_{tt} <0$. Moreover, it admits an $SO(3,\mathbb{R})$ symmetry that will be discussed below. However, as the orbits of the $SO(3,\mathbb{R})$ action are not two-dimensional spacelike spheres, the metric is not spherically symmetric in the usual sense of the word. 

In (\ref{metric}) we have written the metric in a way that involves, in addition to the three parameters $m$, $l$ and $e$ also another, dimensionless, parameter $C$ which was not included in the original work of Brill. It was introduced only later by Manko and Ruiz\cite{MankoRuiz2005} for the NUT metric and it generalizes naturally to the Brill metric. By a coordinate transformation 
\begin{equation}
t'=t-2 l C \varphi,\; r'=r,\;\varphi'=\varphi,\; \vartheta'=\vartheta
\end{equation}
one can transform the Manko-Ruiz parameter $C$ to zero near any one point off the axis, so the local geometry off the axis is unaffected by changing $C$. On the axis, there is a conic singularity, if $l \neq 0$, and this singularity is influenced by $C$: For $C=1$ the singularity is on the upper half axis ($\vartheta =0$), for $C=- 1$ it is on the lower half-axis ($\vartheta = \pi$), and for any other value of $C$ it is on both half-axes, symmetrically distributed for $C=0$ and asymmetrically for other values of $C$. Each of the three parameters $m$, $e$ and $l$ has the dimension of a length. $m$ is the mass parameter which will be assumed non-negative throughout, $m \ge 0$. $e^2= q^2+p^2$ is the combination of an electric charge parameter $q$ and a magnetic charge parameter $p$; obviously $e^2 \ge 0$. $l$ is the gravitomagnetic charge, also known as the NUT parameter, which may take any value $l \in ]-  \infty,+\infty[$. In the analogy between gravitation and electromagnetism, $m$ corresponds to the electric charge, while $l$ corresponds to a magnetic (monopole) charge. 

For $l=0$, the metric reduces to the Reissner-Nordstr{\"o}m metric. Then there is a curvature singularity at $r=0$, so we have to restrict to the region $r>0$ (or to the region $r<0$). If $m>0$, the spacetime region $r>0$ describes a black hole, with horizons at $r_{\pm}= m \pm \sqrt{m^2-e^2}$, for $e^2 \le m^2$ and a naked singularity for $e^2>m^2$; the first case includes of course the Schwarzschild metric with $e^2=0$. If $m=0$, we have a massless naked singularity for $e^2>0$ and flat Minkowski spacetime for $e^2=0$. As the Reissner-Nordstr{\"o}m metric has been extensively covered in the literature, we exclude the case $l = 0$ in the rest of this paper. 

With $l \neq 0$ the Brill metric describes a black hole for $b \le 0$ and a traversable wormhole  for $b>0$, see Cl{\'e}ment et al.  \cite{Clement}. In the case of black holes, there are two horizons at $r_{\pm}=m\pm\sqrt{m^2+l^2-e^2}$. For $b=0$, i.e. $q^2+p^2= m^2+l^2$, it was shown by Cl{\'e}ment et al. \cite{Clement} that the metric is a special case of the Israel-Wilson-Perj\`{e}s metric; in this case we have a black hole with a degenerate horizon. The two special cases $m=0$ and $e^2=0$ are included: $m=0$ gives us a massless black hole for $l^2 \ge e^2$ and a massless wormhole for $l^2<e^2$. $e^2=0$ gives us the NUT metric \cite{NewmanTamburinoUnti1963}. The NUT metric is a solution to Einstein's vacuum field equation that describes a black hole, with horizons at $r_{\pm}=m \pm \sqrt{m^2+l^2}$. The region between the two horizons is isometric to a cosmological vacuum solution found by Taub \cite{Taub1951}; therefore, the analytic extension of the NUT solution beyond the outer horizon is properly called the Taub-NUT solution. 

In the black-hole case, for gravitational lensing it is reasonable to restrict $r$ to the domain of outer communication, i.e., to the region outside of the outer horizon $m+\sqrt{m^2+l^2-e^2}<r<\infty$. Clearly, an observer in the domain of outer communication can receive only light signals that are completely contained in the domain of outer communication, so as long as we do not consider observers who are foolhardy enough to jump into the black hole the region beyond the outer horizon is of no relevance. In the wormhole case, however, there are no horizons which means that any observer can receive signals from the entire domain $- \infty < r < \infty$.

We have said that we assume that $t$ runs over all of $\mathbb{R}$ and that then for $l \neq 0$ there is a conic singularity on the axis. This singularity can actually be removed by making the time coordinate periodic, with the period  $4 \pi | l C|$, as was suggested by Misner \cite{Misner1963} (for the uncharged NUT metric with $C=-1$). This, however, leads to a closed timelike curve through each event where $\partial _t$ is timelike, i.e., to a most drastic kind of causality violation, so we will not follow this suggestion. It is true that also without making the time coordinate periodic there are closed timelike curves in the Brill (or in particular NUT) spacetime, but if the NUT parameter is sufficiently small they are restricted to an arbitrarily small region near the axis, so one may argue that this does not lead to any pathological behaviour that is actually observable.


In the rest of this paper, a Brill spacetime with $m \ge 0$ and $l \neq 0$ will be assumed. For $b \le 0$, we limit ourselves to the domain of outer communication of the black hole, whereas in the wormhole case $b>0$ we have to consider the entire domain $- \infty < r < \infty$. It was emphasized already by Cl{\'e}ment et al. \cite{Clement} that Brill wormholes are traversable, i.e., that luminal and subluminal signals can travel from $r=- \infty$ to $r=\infty$ and vice versa. This distinguishes Brill wormholes from the Einstein-Rosen bridge \cite{EinsteinRosen1935}. As the Brill metric is a solution to the Einstein-Maxwell equations, there is no exotic matter involved. This dinstinguishes the Brill wormholes from the Teo wormholes \cite{Teo:1998dp} (which include the Morris-Thorne wormholes \cite{MorrisThorne1988}) and also from a class of wormholes considered by Halla and Perlick \cite{M2} that is closely related to but not identical with the class of Teo wormholes. All these wormholes are traversable but by Einstein's field equation they have negative energy densities near the throat. The Brill wormholes do not violate any of the energy conditions (weak, strong or dominant). The price we have to pay is in the weaker asymptotic structure: Whereas Teo wormholes have two ends which are asymptotically flat in the sense that the metric approaches the Minkowski metric, the Brill wormhole metrics are asymptotically flat only in the sense that the curvature goes to zero for $r \to \pm \infty$; however, the spheres $(r= \mathrm{const.},t= \mathrm{const.})$ do not become spacelike surfaces with area $4 \pi r^2$ for big $r$. Also, there is the above-mentioned conical singularity on the axis. Nonetheless, some readers may find it attractive to have wormhole solutions to Einstein's field equation without exotic matter, even if they have some other pathologies.  

\section{Geodesics in the Brill metric}\label{sec:geo}
First, we specify the Killing vector fields in order to determine the symmetry and the geodesic equations of the Brill metric. The metric \eqref{metric} has the following four linearly independent Killing vector fields:
\begin{equation}
\xi _0 = \partial _t \, ,
\label{eq:Killing0}
\end{equation} 
\begin{equation}
\xi _1 = - \mathrm{sin} \, \varphi \, \partial _{\vartheta}
- \dfrac{\mathrm{cos} \, \varphi}{\mathrm{sin} \, \vartheta} \Big(
\mathrm{cos} \, \vartheta \, \partial _{\varphi}
+
2l \big( 1 + C \, \mathrm{cos} \, \vartheta \big) \partial _t \Big)
\, ,
\label{eq:Killing1}
\end{equation} 
\begin{equation}
\xi _2  =  \mathrm{cos} \, \varphi \, \partial _{\vartheta}
- \dfrac{\mathrm{sin} \, \varphi}{\mathrm{sin} \, \vartheta} \Big(
\mathrm{cos} \, \vartheta \, \partial _{\varphi}
+
2l \big( 1 + C \, \mathrm{cos} \, \vartheta \big) \partial _t \Big)
\, ,
\label{eq:Killing2}
\end{equation} 
\begin{equation}
\xi _3 = \partial _{\varphi} + 2 l C \partial _t
\, \ .
\label{eq:Killing3}
\end{equation} 
The Killing vector fields satisfy the following Lie bracket relations:
\begin{equation}
\big[ \xi _0 , \xi _1 \big] = 
\big[ \xi _0 , \xi _2 \big] = 
\big[ \xi _0 , \xi _3 \big] = 0 \, ,
\label{eq:Lie1}
\end{equation}
\begin{equation}
\big[ \xi _1 , \xi _2 \big] = - \xi _3 \, , \:
\big[ \xi _2 , \xi _3 \big] = - \xi _ 1 \, , \:
\big[ \xi _3 , \xi _1 \big] = - \xi _2  \, .
\label{eq:Lie2}
\end{equation}
Hence, $\xi _1$, $\xi _2$ and $\xi _3$ generate a three-dimensional group of isometries which is isomorphic to the rotation group $SO(3,\mathbb{R})$, and $\xi_0$ generates a one-dimensional group of isometries that expresses stationarity. Note that for $l \neq 0$ the $SO(3, \mathbb{R} )$ orbits are \emph{not} two-dimensional spacelike spheres but rather three-dimensional submanifolds with topology $\mathbb{S}^2 \times \mathbb{R}$ and signature $(-,+,+)$, so the metric is not spherically symmetric in the usual sense. 

For the NUT metric ($e=0$) with arbitrary Manko-Ruiz parameter $C$, the Killing vector fields have already been given by Halla and Perlick \cite{M1}. For the NUT metric with $C=-1$ they are known from the original NUT paper \cite{NewmanTamburinoUnti1963}. Note that the Killing vector fields and their Lie bracket relations involve neither $m$ nor $e$.

We will now use the Killing symmetries to solve the geodesic equations of the metric \eqref{metric}, where the geodesics $x^{\mu}(s)$ are parametrized by an affine parameter $s$. We denote the derivative with respect to $s$ by an overdot. If we send $l$ to 0 the following analysis gives the well-known geodesics in the Reissner-Nordstr{\"o}m metric, cf. e.g. Chandrasekhar \cite{Chandrasekhar1983}. For $e=0$ it gives the geodesics in the NUT metric which have been discussed by Zimmerman and Shahir \cite{ZimmermanShahir1989} and in even greater mathematical detail by Kagramanova et al. \cite{KagramanovaEtAl2010}.  

Since there are four Killing vector fields $\xi _A$ (where $A\in \{0,1,2,3\}$), there are also four constants of motion $g_{\mu \nu}\xi _A ^{\mu} \dot{x}{}^{\nu}$:
\begin{equation}
E = -g_{\mu \nu} \xi_0 ^{\mu} \dot{x}{}^{\nu} =
\dfrac{r^2+l^2}{r^2-2mr-l^2+e^2} \, 
\Big( 
\dot{t} - 2l \big( \mathrm{cos} \, \vartheta +C \big) \, \dot{\varphi} 
\Big)    
\, ,
\label{eq:E}
\end{equation}

\[
J_1 = g_{\mu \nu} \xi_1 ^{\mu} \dot{x}{}^{\nu} =
- \mathrm{sin} \, \varphi \, (r^2+l^2) \, \dot{\vartheta} 
\]
\begin{equation}
-
\mathrm{cos} \, \varphi \, \mathrm{cos} \, \vartheta \,
\mathrm{sin} \, \vartheta \, (r^2+l^2) \, \dot{\varphi}
+2 \, l \, E \, \mathrm{cos} \, \varphi \, \mathrm{sin} \, \vartheta
\, ,
\label{eq:J1}
\end{equation}

\[
J_2 = g_{\mu \nu} \xi_2 ^{\mu} \dot{x}{}^{\nu} =
\mathrm{cos} \, \varphi \, (r^2+l^2) \, \dot{\vartheta} 
\]
\begin{equation}
-
\mathrm{sin} \, \varphi \, \mathrm{cos} \, \vartheta \,
\mathrm{sin} \, \vartheta \, (r^2+l^2) \, \dot{\varphi}
+2 \, l \, E \, \mathrm{sin} \, \varphi \, \mathrm{sin} \, \vartheta
\, ,
\label{eq:J2}
\end{equation}

\begin{equation}
J_3 = g_{\mu \nu} \xi_3 ^{\mu} \dot{x}{}^{\nu} =
\mathrm{sin}{}^2 \vartheta \, (r^2+l^2) \, \dot{\varphi}
+2 \, l \, E \, \mathrm{cos} \, \vartheta
\, .
\label{eq:J3}
\end{equation}
After a straightforward calculation, one finds that
\begin{equation}
J^2 := J_1^2 +J_2^2+J_3^2 = 
 (r^2+l^2)^2 \, \big( \dot{\vartheta}{}^2 + \mathrm{sin}{}^2 \vartheta \,
 \dot{\varphi}{}^2 \big) + 4 \, l^2 E^2
\label{eq:J}
\end{equation}
and
\begin{equation}
\begin{pmatrix}
\, J_1 \, \\ J_2 \\ J_3
\end{pmatrix}
\boldsymbol{\cdot}
\begin{pmatrix}
\, \mathrm{cos} \, \varphi \, \mathrm{sin} \, \vartheta \, 
\\ 
\, \mathrm{sin} \, \varphi \, \mathrm{sin} \, \vartheta \, \\ 
\mathrm{cos} \, \vartheta 
\end{pmatrix}
=
2 \, l \, E 
\label{eq:cone}
\end{equation}
where the dot denotes the usual scalar product in Euclidean 3-space. This equation says that every geodesic lies in a (coordinate) cone whose symmetry axis is spanned by the  vector $(J_1,J_2,J_3)$. In the Reissner-Nordstr{\"o}m case $l =0$ this cone becomes a plane, for all geodesics. In Brill spacetimes with $l \neq 0$ it becomes a plane only for geodesics with $E =0$. We will see below that geodesics with $E=0$ are necessarily spacelike. 

The Lagrangian 
\begin{equation}
\mathcal{L} = 
\dfrac{1}{2} \, g_{\mu \nu} \dot{x}{}^{\mu} \, \dot{x}{}^{\nu}
\label{eq:L}
\end{equation}
is an additional constant of motion. Using the constants of motion $E$, $J^2$, $J_3$ and $\mathcal{L}$, the geodesic equations can be written in first-order form. For that purpose, one has to solve Eqs. (\ref{eq:J3}), (\ref{eq:J1}), (\ref{eq:J}) and (\ref{eq:L}) in this order for $\dot{\varphi}$, $\dot{t}$, $\dot{\vartheta}{}^2$ and $\dot{r}{}^2$. One gets 
\begin{equation}
\dot{\varphi} = 
\dfrac{
J_3-2 \, l \, E\, \mathrm{cos} \, \vartheta
}{
(r^2+l^2) \, \mathrm{sin} ^2 \vartheta
}
\, ,
\label{eq:geo1}
\end{equation}
       
\begin{equation}
\dot{t} = \dfrac{(r^2-2mr-l^2+e^2) \, E}{r^2+l^2}
+ 2 \, l \, \big( \mathrm{cos} \, \vartheta + C \big)
\dfrac{
\big( J_3-2 \, l \, E \, \mathrm{cos} \, \vartheta \big)
}{
(r^2+l^2) \, \mathrm{sin} ^2 \vartheta
}
\, ,
\label{eq:geo2}
\end{equation}

\begin{equation}
\dot{\vartheta}{}^2 = 
\dfrac{
J^2 - 4 \, l^2 E^2}{(r^2+l^2)^2}
-
\dfrac{ 
( J_3-2 \, l \, E\, \mathrm{cos} \, \vartheta )^2
}{
(r^2+l^2)^2 \, \mathrm{sin} ^2 \vartheta
}
\, ,
\label{eq:geo3}
\end{equation}

\begin{equation}
\dot{r}{}^2 = 
E^2 - \dfrac{(r^2-2mr-l^2+e^2)}{(r^2+l^2)^2} \, 
\big( J^2- 4 \, l^2 E^2 \big) 
+ \dfrac{2 \, \mathcal{L} \big( r^2 - 2 m r -l^2+e^2 \big)}{(r^2+l^2)}
\, .
\label{eq:geo4}
\end{equation}
The same set of equations can be derived in the Hamiltonian formalism. Then  one sees that the Hamilton-Jacobi equation separates and that $J^2-4\,l^2E^2$ is the separation constant, commonly known as the Carter constant.

\section{Lightlike geodesics contained in a cone}\label{sec:light}
We will now consider lightlike geodesics, i.e. geodesics with $\mathcal{L}=0$. We have already noticed that, by (\ref{eq:cone}), any geodesic lies in a cone and that this cone becomes a plane, in a Brill spacetime with $l \neq 0$, only if $E=0$. We will now show that $E=0$ is impossible for lightlike, and also for timelike, geodesics. To that end we observe that (\ref{eq:geo4}), with $\mathcal{L} \le 0$ and $E=0$, can hold only with $\dot{r} =0$ and $J=0$. (Recall that in the case of a black-hole spacetime, $b\le 0$, we restrict to the domain of outer communication whereas in the case of a wormhole spacetime, $b>0$, we allow $r$ to take all real values.) As $J=0$ implies $J_3=0$, we can read from (\ref{eq:geo1}), (\ref{eq:geo2}) and (\ref{eq:geo3}) that then $\dot{\varphi}=0$, $\dot{t}=0$ and $\dot{\vartheta} =0$ which means that we do not get a (geodesic) curve as the solution but just a point. 

For calculating the lightlike geodesics it suffices to consider geodesics with $J_1=J_2=0$. By (\ref{eq:cone}), they lie in a cone of the form $\vartheta = \mathrm{constant}$, i.e., in a cone that is symmetric with respect to the vertical coordinate axis. All other lightlike geodesics are then obtained by applying all possible rotations by using the $SO(3, \mathbb{R})$ action.

If $\vartheta = \mathrm{constant}$ and $J_1 = J_2 =0$, Eqs. \eqref{eq:J1} and \eqref{eq:J2} require 
\begin{equation}
\dot{\varphi} = \dfrac{2 \, l \, E}{(r^2+l^2) \, \mathrm{cos} \, \vartheta} \ ,
\label{eq:dphitheta}
\end{equation}
\\
and Eq. (\ref{eq:cone}) yields
\begin{equation}  
\dfrac{J_3}{E}  = \dfrac{2 \, l}{\mathrm{cos} \, \vartheta} \ .
\label{eq:J3theta}
\end{equation}
Eq. \eqref{eq:J3theta} demonstrates that in a Brill  spacetime with $l \neq 0$ the opening angle $\vartheta$ of the cone is determined by the constant of motion $J_3/E$ and vice versa. Such a relation does not exist in the Reissner-Nordstr{\"o}m spacetime, $l=0$, where the opening angle is always $\pi/2$, for all values of $J_3/E$. Eq. \eqref{eq:J3theta} also demonstrates that in a Brill spacetime with $l \neq 0$ the value $\vartheta = \pi /2$ is not allowed, because $J_3/E$ must be finite. So in a Brill spacetime with $l \neq 0$ the only light rays that lie in a (coordinate) plane through the origin are radial ones; such a light ray is allowed because it lies at the same time in a certain cone with opening angle $< \pi /2.$ 

If we insert Eq. (\ref{eq:J3theta}), together with $J_1=J_2=0$, into Eq. (\ref{eq:geo4}) with
$\mathcal{L}=0$, we get
\begin{equation}
\dot{r}{}^2 = E^2 \Bigg( 1 - \dfrac{4 \, l^2 (r^2-2mr-l^2+e^2)}{(r^2+l^2)^2}
\, \mathrm{tan}{}^2 \vartheta \Bigg) \, .
\label{eq:drtheta}
\end{equation}
After differentiation of this equation with respect to the affine parameter $s$ and dividing the resulting equation by $2\dot{r}$, one obtains
\begin{equation}
\ddot{r} =  \dfrac{4 \, l^2 E^2 \, \mathrm{tan}{}^2 \vartheta}{(r^2+l^2)^3}
\Big( r^3 - 3mr^2-3l^2r+2 e^2\; r+m\;l^2 \Big) \, .
\label{eq:ddr}
\end{equation}
By continuity, this equation is valid also at points where $\dot{r}=0$, although we divided by $\dot{r}$. The orbit equation that determines the shape of a lightlike geodesic in the cone $\vartheta = \mathrm{constant}$ can be found by dividing Eq. \eqref{eq:drtheta} by Eq. \eqref{eq:dphitheta}. The result is  
\begin{equation}
\dfrac{dr}{d \varphi} = \dfrac{\dot{r}}{\dot{\varphi}} 
=
\pm  \, \sqrt{ \dfrac{(r^2+l^2) ^2}{4 \, l^2} \, \mathrm{cos}{}^2 \vartheta 
 - ((r-m)^2+b) \, \mathrm{sin}{}^2 \vartheta }
 \, .
\label{eq:orbit}
\end{equation}
It is often convenient to rewrite the orbit equation (\ref{eq:orbit}) in the form
\begin{equation}
 \frac{4 l^2}{\mathrm{sin}^2 \vartheta} \Big( \frac{dr}{d\varphi} \Big)^2+
 V_{\vartheta } (r)=-4 l^2 b \ ,
\end{equation}
where 
\begin{equation}
 V_{\vartheta } (r)=-\dfrac{(r^2+l^2)^2}{\mathrm{tan}^2 \vartheta}+4 l^2(r-m)^2
\label{potV}
\end{equation}
is an effective potential that depends parametrically on $\vartheta$. Obviously, the condition $V_{\vartheta}(r)\le -4l^2b$ determines the region in the $r-\vartheta-$plane where light rays can exist. 

We now determine the location of \emph{photon circles}, i.e. circular lightlike geodesics, in the Brill spacetime which are of particular relevance for the lensing features. We first consider photon circles about the vertical coordinate axis, i.e. photon circles that occur at the intersection of a sphere $r=r_{\mathrm{ph}}$ with a cone $\vartheta = \vartheta _{\mathrm{ph}}$. To determine $r_{\mathrm{ph}}$ and $\vartheta _{\mathrm{ph}}$ we use Eqs. (\ref{eq:drtheta}) and (\ref{eq:ddr}) for solving the equations $\dot{r} =0$ and $\ddot{r}=0$ simultaneously, which results in
\begin{equation}
r _{\mathrm{ph}}^3 - 3mr _{\mathrm{ph}}^2-3l^2r _{\mathrm{ph}}+2e^2\;r _{\mathrm{ph}}+m\;l^2 
= 0
\label{eq:rph}
\end{equation}
and
\begin{equation}
\mathrm{tan} \, \vartheta _{\mathrm{ph}}= 
\dfrac{r _{\mathrm{ph}}^2+l^2}{2 \, l \, \sqrt{r _{\mathrm{ph}}^2-2mr_{\mathrm{ph}}-l^2+e^2}}
\, .
\label{eq:thetaph}
\end{equation}
With these results at hand, we use Eq. (\ref{eq:geo3}) for solving simultaneously the equations $\dot{\vartheta}=0$ and $\ddot{\vartheta} =0$ which yields
\begin{equation}
    J_3=\dfrac{2lE}{\mathrm{cos} \, \vartheta _{\mathrm{ph}}} 
\end{equation}
and
\begin{equation}
    J^2-4l^2 E^2= 
    \dfrac{(r_{\mathrm{ph}}^2+l^2)^2}{r_{\mathrm{ph}}^2-2mr_{\mathrm{ph}}-l^2+e^2 }
    \, .
\label{eq:Carterph}
\end{equation}
Because of the $SO(3, \mathbb{R})$ symmetry of the spacetime every such photon circle about the vertical coordinate axis gives rise to a \emph{photon sphere} at $r=r_{\mathrm{ph}}$. Through each point of the photon sphere and for each spatial direction tangential to the photon sphere there is a photon circle. Note that in the case $l \neq 0$ a photon circle does not divide the photon sphere into equal halves, i.e., it is not a great circle. Also note that all photon circles in a photon sphere have the same Carter constant, given by Eq. (\ref{eq:Carterph}).   

By construction, a photon circle $(r=\mathrm{const.},\vartheta = \mathrm{const.})$ occurs at those $(r, \vartheta )$ values where $V_{\vartheta} (r) = - 4 l^2b$ and $dV_{\vartheta}(r)/dr=0$. The photon circle is stable if $d^2V_{\vartheta} (r)/dr^2 > 0$ and it is unstable if $d^2V_{\vartheta} (r)/dr^2 < 0$. Here, calling a photon circle ``stable'' means that a small radial perturbation gives a light ray that oscillates about the photon circle, while calling it ``unstable'' means that it gives a light ray that goes away from the photon circle. Clearly, because of the $SO(3,\mathbb{R})$ symmetry we may give the attribute of being stable or unstable to the entire photon sphere. In the $r-\vartheta-$plane photon circles occur at those points where the curve $V_{\vartheta} (r) = - 4 l^2b$ has a horizontal tangent. The corresponding photon sphere is stable if near this point the allowed region $V_{\vartheta}<-4l^2b$ is convex and it is unstable if the forbidden region $V_{\vartheta}>-4l^2b$ is convex. Below we will use the  effective potential $V_{\vartheta}$ for discussing the lensing features and in particular the photon spheres for Brill black holes and Brill wormholes separately.

As Brill metrics are contained in the class of Pleba{\'n}ski metrics, their photon spheres are special cases of photon regions in Pleba{\'n}ski spacetimes which were discussed by Grenzebach et al. \cite{GrenzebachEtAl2015}. We will use their results below for determining the shadow of a Brill black hole and of a Brill wormhole. The relevance of photon spheres, and more generally photon regions, for calculating the shadow of a compact object is detailed in a review by Perlick and Tsupko \cite{PerlickTsupko2022}.

For the NUT metric ($e=0$), the fact that (lightlike) geodesics are contained in a cone is known since quite some time, see Zimmerman and Shahir \cite{ZimmermanShahir1989}. For this case it was shown by Halla and Perlick \cite{M1} that the lightlike geodesics in a cone $\vartheta = \mathrm{constant}$ are geodesics of a two-dimensional Riemannian metric, called the \emph{optical metric}. The generalization of this construction to Brill metrics with $e \neq 0$ is quite straight-forward, so we do not repeat the details here but just give the result: In a general Brill spacetime, the optical metric on a cone $\vartheta = \mathrm{constant}$ reads  
\begin{equation}
\bar{g}{}_{ij}dx^i dx^j=\dfrac{(r^2+l^2)^2}{(r^2-2mr-l^2+e^2)^2}\;dr^2
+\dfrac{(r^2+l^2)^2}{(r^2-2mr-l^2+e^2)} \mathrm{sin}{}^2 \vartheta\; d\varphi^2
\, .
\label{eq:opt}
\end{equation}
Here and in the following the summation convention is used for latin indices that take the values 1 and 2, where $x^1=r$ and $x^2=\varphi$. Note that whereas every lightlike geodesic in the cone $\vartheta = \mathrm{constant}$ is a geodesic of the optical metric, it is of course \emph{not} true that every geodesic of the optical metric is a lightlike geodesic. 

The Gaussian curvature of the metric (\ref{eq:opt}) is 
\begin{equation}
\begin{aligned}
K(r)&= - \dfrac{(2 r-3 m ) m \, r^4}{(l^2+r^2)^4}\\
&-\dfrac{l^2 r^2 \left(14 m^2-20 m r+7 
r^2\right) + l^4 \left(m^2+10 m r-6 r^2\right) + 3 l^6}{(l^2+r^2)^4}\\
&-\dfrac{e^2(r^2-2mr-l^2)(5l^2-3r^2)-2e^4(r^2-l^2)}{(l^2+r^2)^4}
 \, .
\label{eq:K}
\end{aligned}
\end{equation}
This demonstrates that the ``cone'' $\vartheta = \mathrm{constant}$ is a (flat) cone only in the coordinate represention. As $K(r) \neq 0$, the intrinsic geometry of this ``cone'' with the optical metric is not flat. By Eq. (\ref{eq:K}), the Gaussian curvature $K(r)$ is independent of $\vartheta$. This means that, on all cones with their different opening angles, the Gaussian curvature is given by the same function of $r$. The reason for this becomes obvious if we change the coordinates from $(r, \varphi )$ to $( \tilde{r} = r , \tilde{\varphi} = \varphi \, \mathrm{sin} \, \vartheta  )$ in Eq. (\ref{eq:opt}). After this, the metric coefficients are independent of $\vartheta$. Accordingly, the optical metrics of any two cones with different opening angles are locally isometric. However, they are not globally isometric, as the range of the coordinate $\tilde{\varphi}$ depends on $\vartheta $. 

The intrinsic geometry of the two-dimensional Riemannian manifold (\ref{eq:opt}) can be visualized by isometrically embedding it into Euclidean 3-space as a surface of revolution. In cylindrical polar coordinates ($Z$, $R$, $\varphi$), the condition
\begin{equation}
\bar{g}_{ij}dx^idx^i=dZ^2+dR^2+R^2d \varphi ^2 \ ,
\label{optical}
\end{equation}
where $Z(r)$ and $R(r)$ are embedding functions, has to be satisfied. If one inserts Eq. (\ref{eq:opt}) and compares the coefficients of $d \varphi ^2$ and  $dr^2$, the results are
\begin{equation}
R(r) =
\dfrac{(r^2+l^2) \mathrm{sin} \, \vartheta}{\sqrt{r^2-2mr-l^2+e^2}} 
\label{eq:emb0}
\end{equation}
and
\[
\Big(\dfrac{d Z(r)}{dr} \Big)^2
= \dfrac{
(r^2-2 m r-l^2+e^2)(r^2+l^2)^2
}{
(r^2-2m r-l^2+ e^2)^3
}
\]
\begin{equation}
-\dfrac{
\mathrm{sin} ^2 \vartheta (r^3-3m r^2-3l^2 r+2e^2\;r+m\;l^2)^2
}{
(r^2-2m r-l^2+ e^2)^3 
} =: H(r) \ .
\label{eq:emb}
\end{equation}
Note that the right-hand side of Eq. (\ref{eq:emb0}) is real and positive on the domain of outer communication in the black-hole case while it is real and positive everywhere in the wormhole case. So the embedding is possible if and only if the function $H(r)$ defined in (\ref{eq:emb}) is non-negative. Comparison with Eq. (\ref{eq:rph}) shows that this condition is always satisfied near a photon circle.

To show how photon circles are represented in the embedded surface we differentiate (\ref{eq:emb0}) and express the resulting equation with the help of the effective potential from Eq. (\ref{potV}) as
\begin{equation}
    \dfrac{dR(r)}{dr} = \dfrac{\Big(4r\big( V_{\vartheta} (r)+4 \ell^2 b \big)-(r^2+l^2) dV_{\vartheta}(r)/dr \Big) \mathrm{sin} \, \vartheta }{ 8 \ell^2 \sqrt{r^2-2mr-l^2+e^2}^{\, 3}}
    \, .
\end{equation}
At a photon circle we have $V_{\vartheta}(r)=-4l^2b$ and $dV_{\vartheta}(r)/dr=0$, hence
\begin{equation}
    \dfrac{dR(r)}{dr} = 0 \, , \quad \dfrac{d^2R(r)}{dr^2}=\dfrac{-(r^2+l^2) \big( d^2V_{\vartheta}(r)/dr^2 \big)}{8 \ell^2 \sqrt{r^2-2mr-l^2+e^2}^{\, 3} } \, .
\label{eq:phembed}
\end{equation}
These equations demonstrate that an unstable photon circle corresponds to a ``neck'' and a stable photon circle to a ``belly'' of the embedded surface. The same result can also be expressed in terms of the Gaussian curvature which, for any surface embedded into Euclidean 3-space, can be written as the product of two principal curvatures, 
\begin{equation}
    K(r) = K_1(r) K_2(r) \, .
\end{equation}
For a surface of revolution, $K_1(r)$ is always positive; in the case at hand, a straight-forward calculation gives
\begin{equation}
    K_1(r) = \sqrt{
    \dfrac{r^2-2mr-\ell^2+e^2}{(r^2+\ell^2)^2 \mathrm{sin} ^2 \vartheta}
    - 
    \dfrac{(r^3-3mr^2-3\ell^2r+2 e^2 r+m \ell^2)^2}{(r^2+\ell^2)^4}
    }
\end{equation}
which is indeed real and positive on the domain where $H(r) >0$. Therefore, the sign of $K(r)$ is determined by the sign of $K_2(r)$. As, by definition of the principal curvatures, $K_2(r)$ is positive near a ``belly'' and negative near a ``throat'', we see that the Gaussian curvature is negative near an unstable photon circle and positive near a stable photon surface. This observation, which is in agreement with geometric intuition, was already made by Qiao and Li \cite{QiaoLi2022,Qiao2022} for the optical metric of a spherically symmetric and static spacetime. Their argument was based on the assumption that the optical metric is complete; the above reasoning demonstrates that this assumption is actually not necessary.

Far away from the center we can approximate the Gaussian curvature and the inverse of the slope of the embedded surface by a Taylor expansion,
\begin{equation}
    K(r) = \dfrac{1}{r^2} \Big( - \dfrac{2m}{r} + O (2)  \Big) 
\label{eq:TaylorK1}
\end{equation}
and
\begin{equation}
    \Big( \dfrac{dR}{dZ} \Big)^2 = \mathrm{tan} ^2 \vartheta \Big( 1 - \dfrac{4m}{r \mathrm{cos}^2 \vartheta} + O(2) \Big)
\label{eq:TaylorRZ1}
\end{equation}
where $O(n)$ stands for terms of $n$th or higher order with respect to $m/r$, $l/r$ and $e/r$. This demonstrates that the embedded surface asymptotically approaches a (flat) cone of opening angle $\vartheta$ in the ambient Euclidean space. If $m=0$, the expansion up to the next-to-leading term reads
\begin{equation}
    K(r) =  \dfrac{1}{r^2} \Big( - \dfrac{7 l^2-3 e^2}{r^2}  + O (3)  \Big) \, , 
\label{eq:TaylorK2}
\end{equation}
and
\begin{equation}
    \Big( \dfrac{dR}{dZ} \Big)^2 = \mathrm{tan} ^2 \vartheta \Big( 1 - \dfrac{7 l^2-3 e^2}{r^2 \mathrm{cos}^2 \vartheta} + O(3) \Big) \, .
\label{eq:TaylorRZ2}
\end{equation}
If both $m=0$ and $7l^2=3e^2$, we have
\begin{equation}
    K(r) = \dfrac{1}{r^2} \Big( -\dfrac{16 l^4}{9r^4}  + O (5) \Big) 
\label{eq:TaylorK3}
\end{equation}
and
\begin{equation}
    \Big( \dfrac{dR}{dZ} \Big)^2 = \mathrm{tan} ^2 \vartheta \Big( 1 - \dfrac{8l^4}{9 r^4 \mathrm{cos}^2 \vartheta} + O(5) \Big) \, .
\label{eq:TaylorRZ3}
\end{equation}
We see that in all three cases the asymptotic cone is approached from the outside if the Gaussian curvature is negative whereas it is approached from the inside if the Gaussian curvature is positive. This is of course in agreement with geometric intuition. Also, the asymptotic formulas confirm our earlier observation that the global geometry of the surfaces $\vartheta = \mathrm{constant}$ depends on $\vartheta$ although the Gaussian curvature does not. 

We will further discuss the geometry of the surfaces $\vartheta = \mathrm{constant}$ and their embedding into Euclidean 3-space for the black hole case ($b \le 0$) in Section \ref{black} and for the wormhole case ($b>0$) in Section \ref{NUTW}. 
\section{Gravitational lensing of Brill black holes}\label{black}   
Throughout this section we consider a Brill spacetime with $m \ge 0$, $l \neq 0$ and $b \le 0$, and we restrict ourselves to the domain of outer communication , $r>m+\sqrt{-b}$. We will see that the lensing features of a Brill black hole are qualitatively more or less the same as those of a  NUT black hole, i.e. that the charge parameter $e$ does not change much. 

\subsection{Photon sphere and shadow}
In the black-hole case $b \le 0$ considered here, the cubic (\ref{eq:rph}) has exactly one solution $r_{\mathrm{ph}}$ in the domain of outer communication which can be written with the help of the standard trigonometric solution formula as
\begin{equation}
    r_{\mathrm{ph}} = m + \sqrt{\dfrac{4}{3} \Big( m^2 + l^2 - 2 b \Big) } \,
    \mathrm{cos} \Bigg( \dfrac{1}{3} \mathrm{arccos} \Big( \dfrac{- \sqrt{27} m b}{\sqrt{m^2+l^2-2b}^{\,3}} \Big) \Bigg) \, . 
\label{eq:bhrph}
\end{equation}
The opening angle $\vartheta _{\mathrm{ph}}$ of the corresponding cones is given by Eq. (\ref{eq:thetaph}). Note that the expression under the square-root in Eq. (\ref{eq:thetaph}) is strictly positive because $r_{\mathrm{ph}}$ is in the domain of outer communication.

\begin{figure}[H]
\centering
\includegraphics[width=0.45\linewidth]{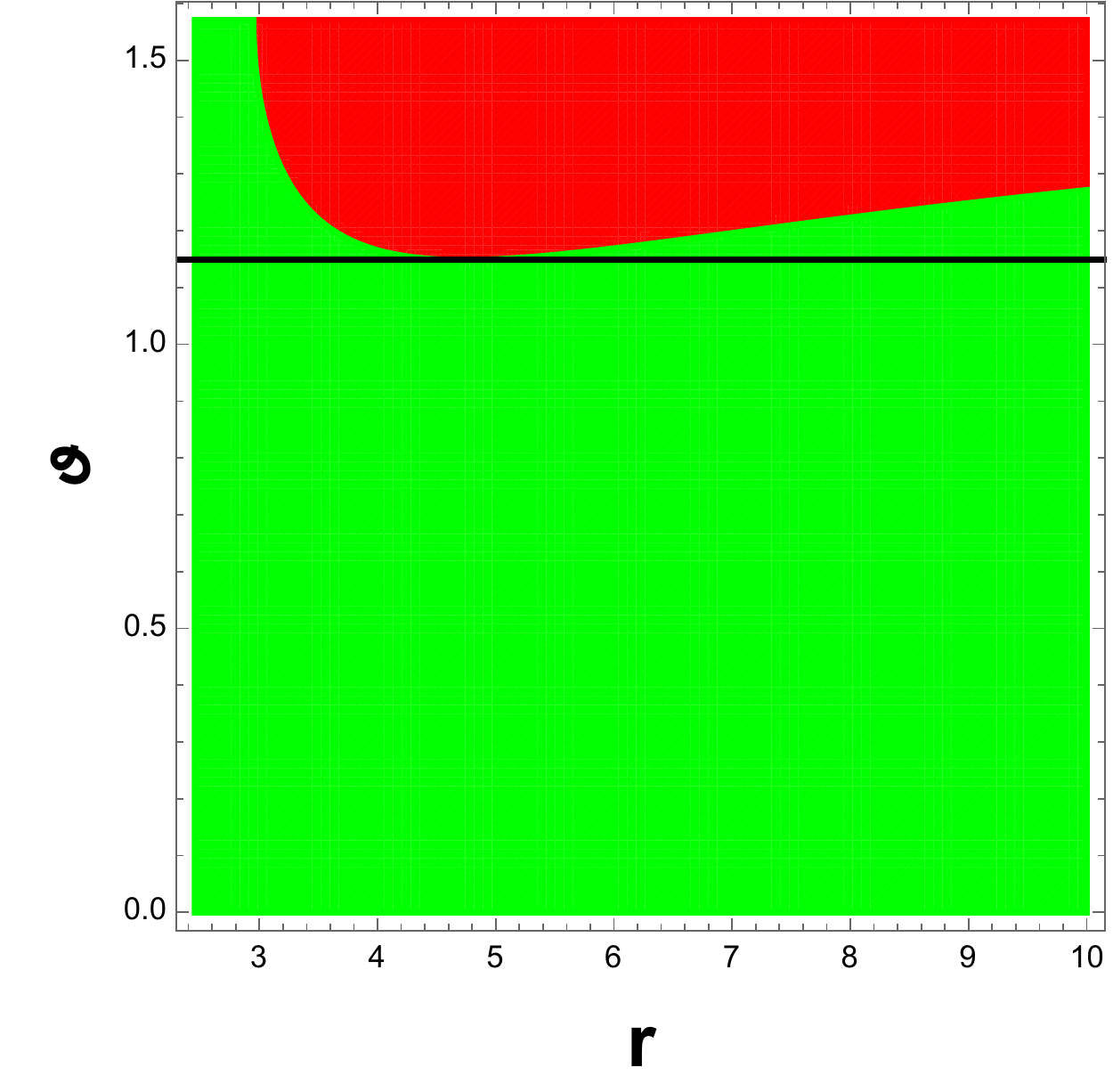}%
\caption{The allowed region $V_{\vartheta}(r) \le - 4 l^2 b$ (green) for light rays propagation in the case of the black hole, with $l=1.81 m$ and $e=0.7 m$. The $r$-axis starts at the horizon $r_{+}=2.45m$ }
  \label{P1_black}
\end{figure}

By differentiating the potential $V_{\vartheta}$ defined in Eq. (\ref{potV}) we find that the photon sphere is unstable. The region allowed for light rays is determined by the condition $V_{\vartheta} (r) \le - 4 l^2 b$. This region is shown, in an $r-\vartheta-$diagram, in green in Fig.~\ref{P1_black}. The forbidden region is shown in red. In this diagram, every light ray moves along a horizontal line in the green region. The black line is the horizontal line at the $\vartheta$ value of the photon sphere $\vartheta _{\mathrm{ph}}$. 

We see that a light ray that comes in from infinity goes to the horizon, without a turning point, if $\vartheta < \vartheta _{\mathrm{ph}}$. However, if $\vartheta > \vartheta _{\mathrm{ph}}$ it goes through a minimum radius $r_m$ when it reaches the boundary of the green region and then goes back to infinity. The borderline cases between these two classes are light rays that asymptotically spiral towards a photon circle at $r=r_{\mathrm{ph}}$ in the cone $\vartheta = \vartheta _{\mathrm{ph}}$. In the next subsection we will calculate the deflection angle which is defined for those light rays that go through a minimum radius. 

Analogously, we read from Fig.~\ref{P1_black} that a light ray that starts just outside the horizon goes to infinity if $\vartheta < \vartheta _{\mathrm{ph}}$, whereas it is reflected and returns to the horizon if $\vartheta > \vartheta _{\mathrm{ph}}$. Again, the borderline cases are light rays that asymptotically spiral towards a photon circle in the photon sphere, now from below. 

\begin{figure}[H]
\centering
    \includegraphics[width=0.8\textwidth]{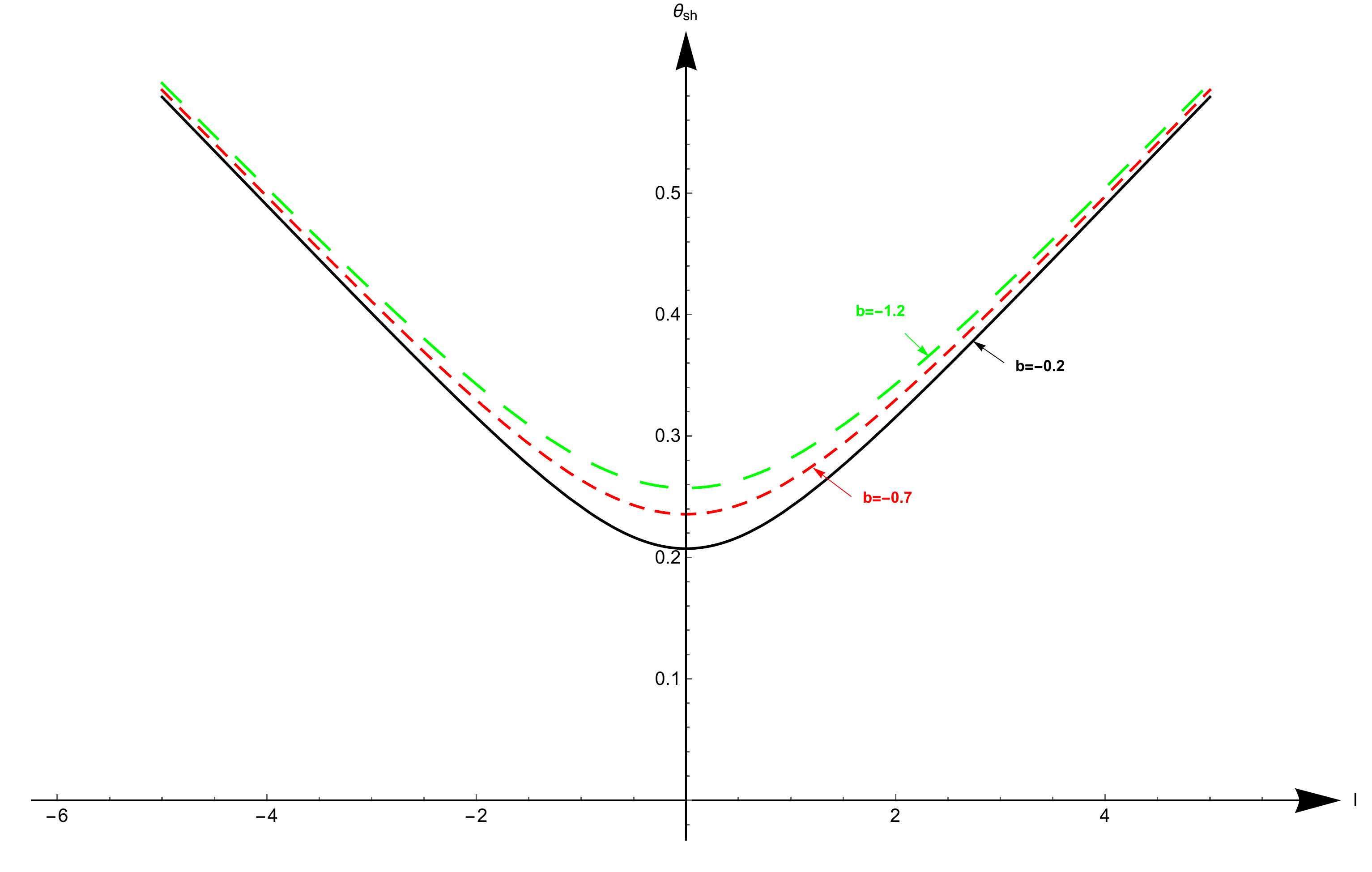}
    \caption{Shadow radius of a Brill black hole, for an observer at $r_{\mathrm{O}}=20m$, with $l$ and $b$ given in units with $m=1$.}
    \label{fig:sh1}
\end{figure}

With Eq. (\ref{eq:bhrph}) at hand we can now determine the shadow of a Brill black hole. If we consider a stationary observer, i.e. an observer at fixed coordinates $(r_{\mathrm{O}},\vartheta _{\mathrm{O}},\varphi _{\mathrm{O}})$, in the domain of outer communication, we may divide all past-oriented light rays issuing from the observer position into two classes: Those which go to infinity and those which go to the horizon. Again, the borderline cases between these two classes are light rays that spiral towards the photon sphere. If we assume that there are no light sources in the region bounded by past-oriented light rays that spiral from the observer position towards the photon sphere, the observer will see a black disk in the sky whose boundary curve is determined by the initial directions of light rays that spiral towards the photon sphere. In the case of a Brill black hole, the shadow is circular and its angular radius $\theta _{\mathrm{sh}}$ is given by specializing Eq. (24) of Grenzebach et al \cite{GrenzebachEtAl2015} to the case at hand. With the Carter constant from Eq. (\ref{eq:Carterph}) this results in   
\begin{equation}
\mathrm{sin} \, \theta _{\mathrm{sh}} = 
\dfrac{(r_{\mathrm{ph}}^2+l^2)}{(r_{\mathrm{O}}^2+l^2)}
\, 
\sqrt{ \dfrac{r_{\mathrm{O}}^2-2mr_{\mathrm{O}}-l^2+e^2}{r_{\mathrm{ph}}^2-2mr_{\mathrm{ph}}-l^2+e^2}}
\, . 
\label{eq:shadowbh} 
\end{equation}
If $r_{\mathrm{O}}$ varies from $r_+$ to infinity, $\theta _{\mathrm{sh}}$ decreases monotonically from $\pi$ to 0, i.e., if the observer position approaches the horizon the entire sky becomes dark and if it approaches infinity the entire sky becomes bright. At $r_{\mathrm{O}}=r_{\mathrm{ph}}$ we have $\theta _{\mathrm{sh}}= \pi/2$, i.e., half of the sky is dark. The depence of the shadow radius on the parameters $l$ and $b$ is shown, for a fixed radius coordinate of the observer, in Fig.~\ref{fig:sh1}. Keep in mind that Eq. (\ref{eq:shadowbh}) is valid only for a stationary observer. For a non-stationary observer we have to apply the aberration formula to Eq. (\ref{eq:shadowbh}). As the relative velocity of the observer with respect to stationary observers will not in general be constant, the shadow will depend on time. Note, however, that it will always be circular because the aberration formula maps circles in the sky onto circles in the sky.

\subsection{Deflection angle}\label{sec:defl}
\begin{figure}[H]
\centering
    \includegraphics[width=0.55\textwidth]{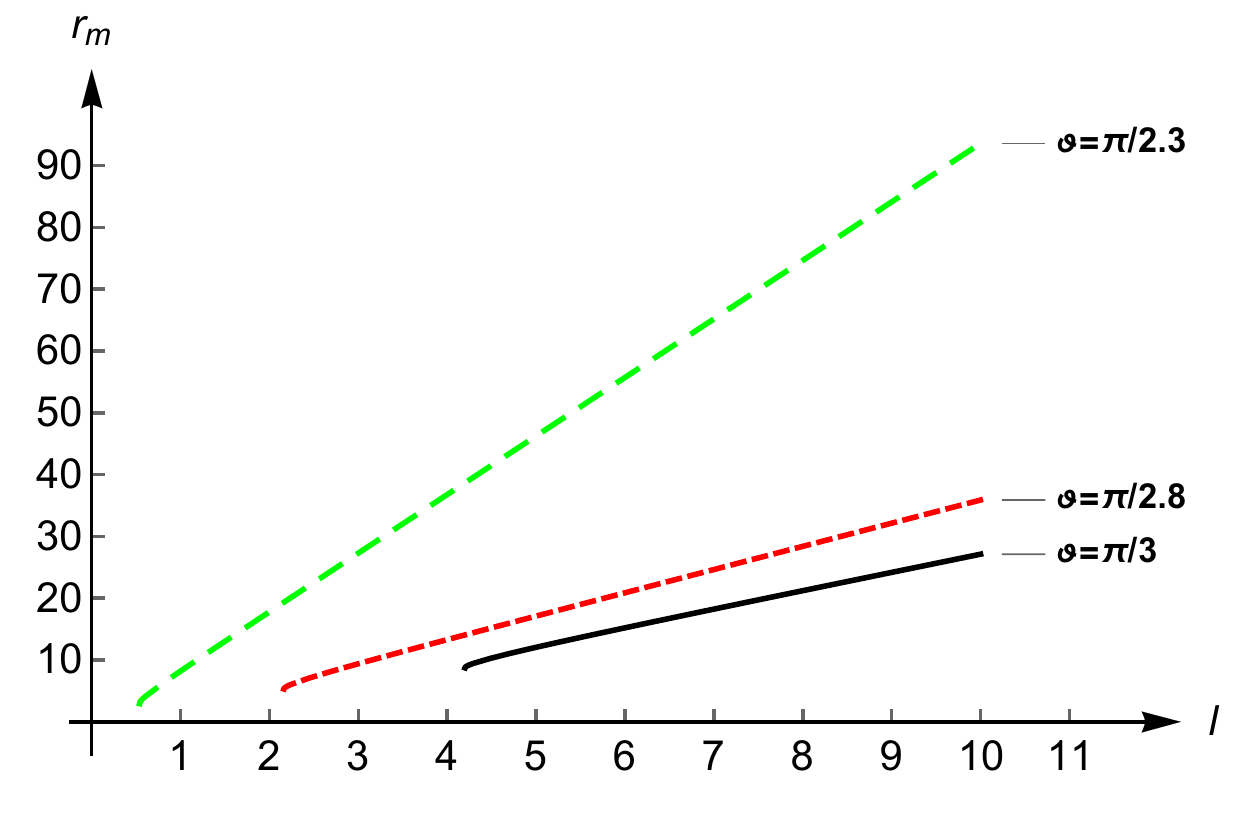}
    \caption{The minimal radius $r_m$ as a function of $l$ for various
    values of $\vartheta$ and $e=0.5$, in units with $m=1$.}
\label{rmf1}
\end{figure}

In order to calculate the deflection angle, we consider a light ray in the cone $\vartheta=$ constant that comes in from infinity, goes through a minimum radius value at $r = r_m$ and then escapes back to infinity. Notice that such a lightlike geodesic can make any number of turns around the center. From Eq. (\ref{eq:drtheta}), where the right-hand side has to be zero at $r=r_m$, one finds
\begin{equation}
\dfrac{r_m^2+l^2}{\sqrt{r_m^2-2 m r_m -l^2+e^2}}= 2 \, l \, \tan\vartheta \, .
\label{eq:rm}
\end{equation}
For given $\vartheta$, this is a fourth-order equation for $r_m$. We have to choose the largest solution $r_m>r_+$. Note that the right-hand side of Eq.~(\ref{eq:ddr}) should be negative because our light rays come in from infinity. Accordingly, light rays that go through a minimal radius value $r_m$ exist for all $r_m$ outside of the photon sphere, $r_m > r_{\mathrm{ph}}$, and the corresponding values of $\vartheta$ converge towards $\vartheta_{\mathrm{ph}}$ for $r_m \to r_{\mathrm{ph}}$. Also note that Eq. (\ref{eq:rm}) can be rewritten with the effective potential from Eq. (\ref{potV}) as $V_{\vartheta} (r_m) =-4 l^2b$. In Fig. \ref{rmf1}, the minimal radius $r_m$ as a function of $l$ is shown for a fixed value of $e^2$ and different opening angles $\vartheta$.

By inserting Eq. (\ref{eq:rm}) into Eq. (\ref{eq:orbit}), the orbit equation can be rewritten as 
\begin{equation}
\dfrac{dr}{d\varphi} 
= 
\pm \, \sqrt{\dfrac{
(r^2+l^2)^2(r_m^2-2 m r_m -l^2+e^2)-(r_m^2+l^2)^2(r^2-2 m r-l^2+e^2)
}{
(r_m^2+l^2)^2+4l^2(r_m^2-2 m r_m-l^2+e^2)
}}
\, .
\label{eq:orbitrm}
\end{equation}
Integrating this equation over the light ray from the point of its closest approach to infinity gives
\begin{equation}
\Delta \varphi = \int_{r_m}^{\infty} 
\dfrac{
\sqrt{(r_m^2+l^2)^2+4l^2(r_m^2-2 m r_m-l^2+e^2)} \, dr
}{
\sqrt{(r^2+l^2)^2(r_m^2-2 m r_m -l^2+e^2)-(r_m^2+l^2)^2(r^2-2 m r-l^2+e^2)}
}
\, .
\label{33}
\end{equation}

\begin{figure}[H]
\centering
    \includegraphics[width=0.65\textwidth]{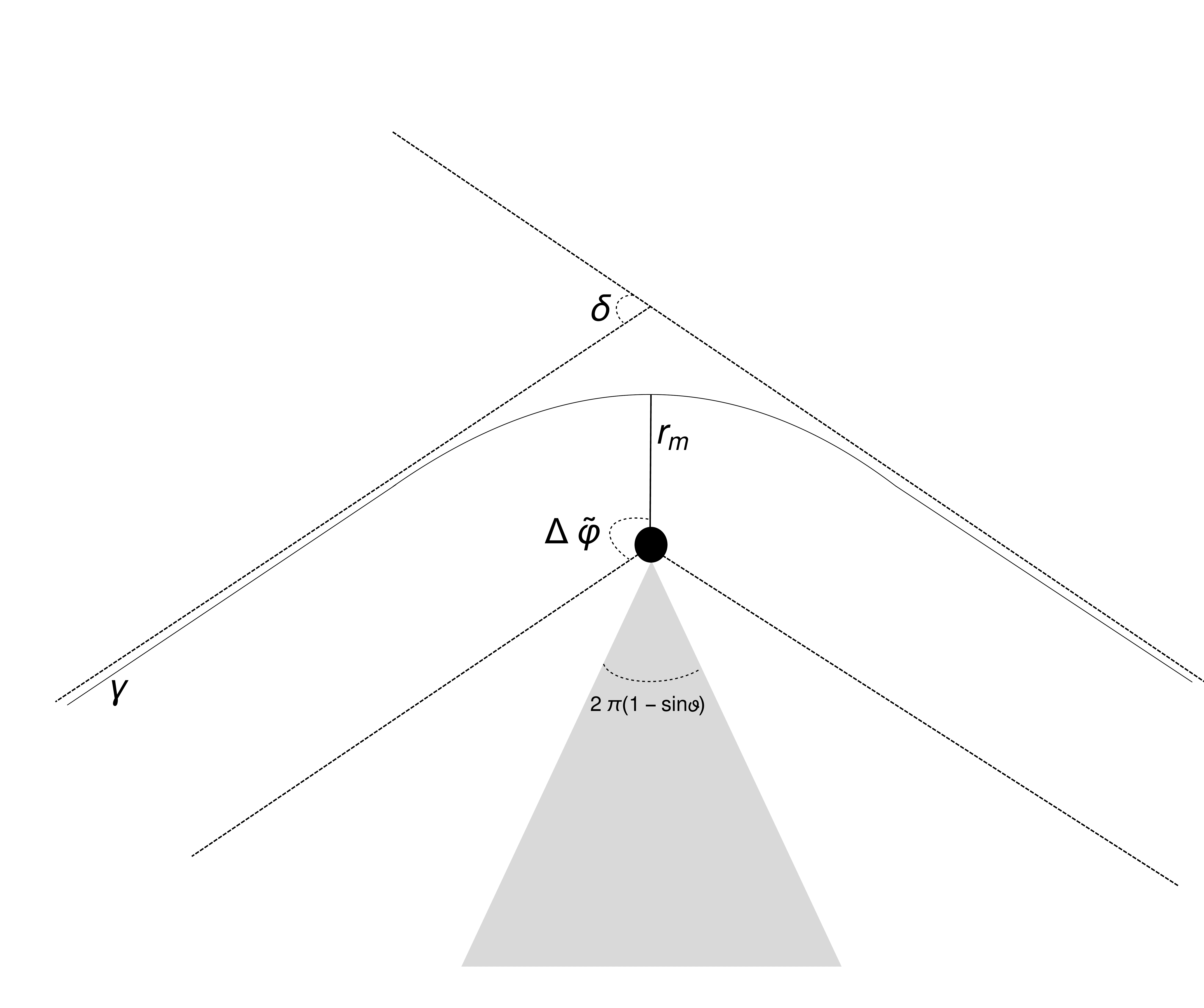}
    \caption{The definition of the deflection angle $\delta$. This picture is reproduced here from Halla and Perlick \cite{M1}.}
\label{rm}
\end{figure}

For determining the deflection angle of a light ray, it is helpful to introduce a new azimuthal coordinate 
\begin{equation}
\tilde{\varphi} = \varphi \, \mathrm{sin} \, \vartheta =
\dfrac{(r_m^2+l^2) \, \varphi}{\sqrt{(r_m^2+l^2)^2
+4 l^2 (r_m^2-2 m r_m-l^2+e^2)}} \ .
\end{equation}
This angle coordinate, which is determined around the surface of the cone, can be visualized by cutting the cone open and flattening it. On each circle $r= \mathrm{constant}$ on the cone, $\tilde{\varphi}$ runs from 0 to $2 \pi \, \mathrm{sin} \, \vartheta$, because $\varphi$ runs from 0 to $2 \pi$, see Fig.~\ref{rm}. As shown in this figure, the deflection angle $\delta$ is defined as the angle under which the asymptotes of the lightlike geodesic intersect in the cut and flattened cone $\vartheta=$ constant,
\begin{equation}
\delta= 2\Delta \tilde{\varphi} -\pi \ ,
\label{def}
\end{equation}
where
\begin{equation}
\begin{aligned}
\Delta \tilde{\varphi} &= \mathrm{sin} \, \vartheta \, \Delta \varphi\\
&=
\int_{r_m}^{\infty} 
\dfrac{
(r_m^2+l^2) \, dr
}{
\sqrt{(r^2+l^2)^2(r_m^2-2 m r_m -l^2+e^2)-(r_m^2+l^2)^2(r^2-2 m r-l^2+e^2)}
}
\, .
\label{eq:Dtphi}
\end{aligned}
\end{equation}

For $r_m \to r_{\mathrm{ph}}$ the deflection angle $\delta$ goes to infinity which means that 
the light ray runs around the center many times. 
Bozza \cite{Bozza2002} has developed an approximation formalism, now known as the \emph{strong deflection limit}, that characterizes the behavior of the deflection angle if $r_m$ comes close to $r_{\mathrm{ph}}$. In his work he assumed that the metric under consideration is spherically symmetric, static and asymptotic Minkowskian. Although the Brill metric shares none of these three properties, Bozza's method can be used for evaluating the deflection angle in this case as well, as we will demonstrate now. To that end we substitute the integration variable $r$ in (\ref{eq:Dtphi}) by a new variable $z$, defined by
\begin{equation}
    z = \dfrac{y(r)-y(r_m)}{1-y(r_m)} \, , \quad y(r) = \dfrac{r^2-2mr-l^2+q^2}{r^2+l^2} \, .
\label{eq:yz}
\end{equation}
Then (\ref{eq:Dtphi}) becomes
\begin{equation}
    \Delta \tilde{\varphi} = \int _0 ^1 R(z,r_m) f(z,r_m) dz
\label{eq:IntRf}
\end{equation} 
with
\begin{equation}
    R(z,r_m) = \dfrac{\big( 1-y(r_m) \big) \sqrt{r_m^2+l^2}}{(r^2+l^2) y'(r)}
    \, , 
\label{eq:R}
\end{equation}
\begin{equation}
    f(z,r_m) = \dfrac{1}{\sqrt{y(r_m)-y(r) \dfrac{(r_m^2+l^2)}{(r^2+l^2)}}} \, .
\label{eq:f}
\end{equation}
In (\ref{eq:R}) and (\ref{eq:f}), $r$ has to expressed in terms of $z$ and $r_m$ with the help of (\ref{eq:yz}).
By Taylor expanding the expression under the square-root up to second order with respect to $z$, $f(z,r_m)$ can be approximated, for small $z$, by
\begin{equation}
    f_0(z,r_m) = \dfrac{1}{\sqrt{\alpha (r_m) z + \beta (r_m) z^2}}
\end{equation}
where
\begin{equation}
    \alpha (r_m) = 1 - y(r_m) \Big( 1 + 
    \dfrac{2 r_m \big( 1 - y(r_m) \big) }{(r_m^2 + l^2) \, y'(r_m)} \Big)
    \, ,
\end{equation}
\[
\beta ( r_m ) = 
    \dfrac{
     2 r_m (r_m^2 + l^2) \big(1 - y(r_m) \big) ^2  
    }{
    (r_m^2 + l^2)^2 y'(r_m)
    }
    \]
\begin{equation}
    -
    \dfrac{
    \big(1 - y(r_m) \big) ^2  y(r_m) 
    }{
    (r_m^2 + l^2)^2 
    }
    \Big(   
    \dfrac{3 r_m^2 - l^2 }{y'(r_m)^2}  
    +
    \dfrac{r_m (r_m^2 + l^2) y''(r_m) }{ y'(r_m)^3}
    \Big)
\end{equation}
Following Bozza's methodology, we now decompose the integral on the right-hand side of (\ref{eq:IntRf}) into
a part 
\begin{equation}
  I_D(r_m) = \int _0 ^1 R(0,r_m) f_0(z,r_m) \, dz
\end{equation}
which diverges for $r_m \to r_{\mathrm{ph}}$ and a part
\begin{equation}
  I_R(r_m) = \int _0 ^1 \big( R(z,r_m) f(z,r_m)-R(0,r_m) f_0(z,r_m)) \, dz
\end{equation}
which is regular for $r_m \to r_{\mathrm{ph}}$. The integral $I_D (r_m)$ can be explicitly calculated and then expanded with respect to $r_m$ about $r_{\mathrm{ph}}$.
\[
    I_D (r_m) = \dfrac{
    2 \big( 1 - y(r_m) \big) \mathrm{log} \Big( \sqrt{\frac{\beta (r_m)}{\alpha (r_m)}}+\sqrt{1+\frac{\beta (r_m)}{\alpha (r_m)}} \Big) 
    }{
    \sqrt{r_m^2+l^2} \, y'(r_m) \, \sqrt{\beta (r_m)}
    }
\]
\begin{equation}
=
- \, \dfrac{\big( 1- y(r_{\mathrm{ph}}) \big) \, \mathrm{log(2)} \Big( \mathrm{log}(r_m-r_{\mathrm{ph}}) - \mathrm{log} \Big(\frac{\beta ( r_{\mathrm{ph}})}{\alpha ' (r_{\mathrm{ph}})} \Big) \Big)
}{
\sqrt{r_{\mathrm{ph}}^2+l^2} \, y'(r_{\mathrm{ph}}) \, \sqrt{\beta (r_{\mathrm{ph}})}
} + O( r_m-r_{\mathrm{ph}})
\end{equation}
where we have used that $\alpha (r_{\mathrm{ph}}) =0$.
The regular integral is of the form
\begin{equation}
    I_R (r_m) = I_R (r_{\mathrm{ph}} ) + O ( r_m-r_{\mathrm{ph}})
\end{equation}
where $I_R (r_{\mathrm{ph}} )$ can be numerically determined for each choice of the parameters $m$, $l$ and $e$. (In the Schwarzschild case, $l=0$ and $e=0$, the integral $I_R (r_{\mathrm{ph}} )$ can be calculated analytically; however, already in the Reissner-Nordstr{\"o}m case, $l=0$ and $e \neq 0$, this is not possible, see Bozza \cite{Bozza2002}.) So we see that in the Brill spacetime the deflection angle diverges logarithmically for $r_m \to r_{\mathrm{ph}}$
\begin{equation}
    \Delta \tilde{\varphi} = I_D ( r_m) + I_R (r_m) =
    - \, \dfrac{\big( 1- y(r_{\mathrm{ph}}) \big) \, \mathrm{log(2)}  
    }{
    \sqrt{r_{\mathrm{ph}}^2+l^2} \, y'(r_{\mathrm{ph}}) \, \sqrt{\beta (r_{\mathrm{ph}})}
    } \, \mathrm{log}(r_m-r_{\mathrm{ph}}) \, + B + O( r_m-r_{\mathrm{ph}})
\end{equation}
with a constant $B$ that has to be determined numerically. As the Brill spacetime doesn't satisfy the assumptions on which Bozza's results were based, this logarithmic behavior could not have been anticipated.

For large $r_m$, on the other hand, $\delta$ is small and may be well aproximated by a low-order Taylor expansion with respect to the dimensionless parameters $m/r_m$, $e/r_m$ and $l/r_m$. As $e$ and $l$ enter quadratically in the metric, one has to go at least up to the second order to have a non-vanishing contribution from them. To within this order, Eq. (\ref{eq:Dtphi}) reduces to
\[
\Delta \tilde{\varphi} 
=
\int_{r_m}^{\infty} 
\left(
1 \, + \, \dfrac{(r^2+r _m r +r_m^2)}{r (r+r_m)} \, \dfrac{m}{r_m}
\right.
\]
\[
+ \, \dfrac{3 (r^2+r_mr+r_m^2)^2}{2 r^2 (r+r_m)^2} \, \dfrac{m^2}{r_m^2}
+ \, \dfrac{(3r^2+r_m^2)}{2 r^2} \, \dfrac{l^2}{r_m^2}
\]
\[
-\dfrac{e^2(r^2+r_m^2)}{2 r^2 r_m^2}\Bigg) \dfrac{r_m \, dr}{r \sqrt{r^2-r_m^2}}  + O(3)
\]
\begin{equation}
=
\dfrac{\pi}{2} + \dfrac{2m}{r_m} - \left( 2 - \dfrac{15 \pi}{8} \right) \dfrac{m^2}{r_m^2}
+ \dfrac{\pi}{8} \, \dfrac{(7l^2-3e^2)}{r_m^2} 
+ O (3) \ ,
\label{eq:Deltatphi}
\end{equation}
where $O(3)$ stands for terms of third or higher order in $m/r_m$, $l/r_m$ and $e/r_m$. Correspondingly, the deflection angle becomes
\begin{equation}
\begin{aligned}
\delta &= 2 \Delta \tilde{\varphi}-\pi\\
&=\dfrac{4m}{r_m} - \left( 4 - \dfrac{15 \pi}{4} \right) \dfrac{m^2}{r_m^2}
+ \dfrac{\pi}{4} \, \dfrac{(7l^2-3e^2)}{r_m^2} 
+ O (3) 
\, .
\end{aligned}
\label{eq:delta}
\end{equation}
For $l = 0$, $\delta$ reduces of course to the well-known expression for the Reissner-Nordstr{\"o}m case.

\begin{figure}[H]
\centering
    \includegraphics[width=0.7\textwidth]{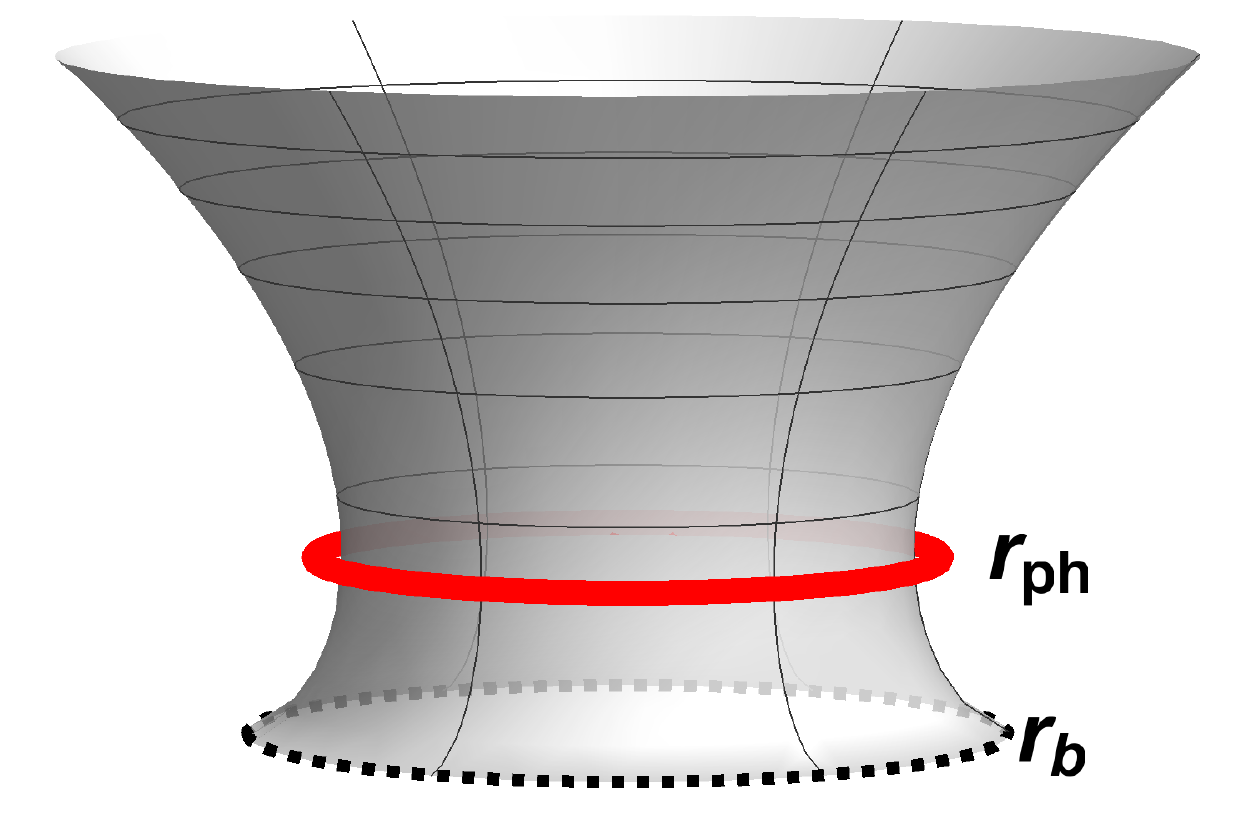}
    \caption{The cone $\vartheta=\vartheta _{\mathrm{ph}}=1.328$ of the optical metric of the Brill metric (\ref{eq:opt}), with $l=0.7m$ and $e=0.5m$, embedded as a surface of revolution into Euclidean 3-space. The photon circle (red) is at $r_{\mathrm{ph}}=3.251 m$ and the boundary of the embeddable part is at $r_{b}= 2.39m$.}
\label{em1}
\end{figure}

We now discuss the intrinsic geometry of the (coordinate) cone $\vartheta = \mathrm{constant}$ with the optical metric (\ref{eq:opt}). An isometric embedding into Euclidean 3-space is possible if the function $H(r)$ defined in (\ref{eq:emb}) is non-negative. We find that, in general, this is true only on an interval $r_b < r < \infty$ with some cutoff-radius $r_b > r_+$ which is determined by the equation $H(r)=0$. As the denominator of $H(r)$ is non-zero on the domain of outer communication, this gives us a sixth-order equation for $r$ which has to be solved numerically. $r_b$ is the greatest zero with $r_+<r_b<\infty$. An example is shown in Fig. \ref{em1}.
We know already from Eqs. (\ref{eq:TaylorK1}) to (\ref{eq:TaylorRZ3}) that the embedded surface approaches a flat cone in the ambient Euclidean space for $r \to \infty$. With our assumptions $l \neq 0$ and $b \le 0$ these equations imply that the Gaussian curvature is always negative near $r = \infty$ and that, correspondingly, the asymptotic cone is approached from the outside. We also know from Eq. (\ref{eq:phembed}) that an unstable photon circle corresponds to a ``neck'' of the embedded surface. These features are illustrated by Fig.~\ref{em1}. 

\begin{figure}[H]
\centering
    \includegraphics[width=0.7\textwidth]{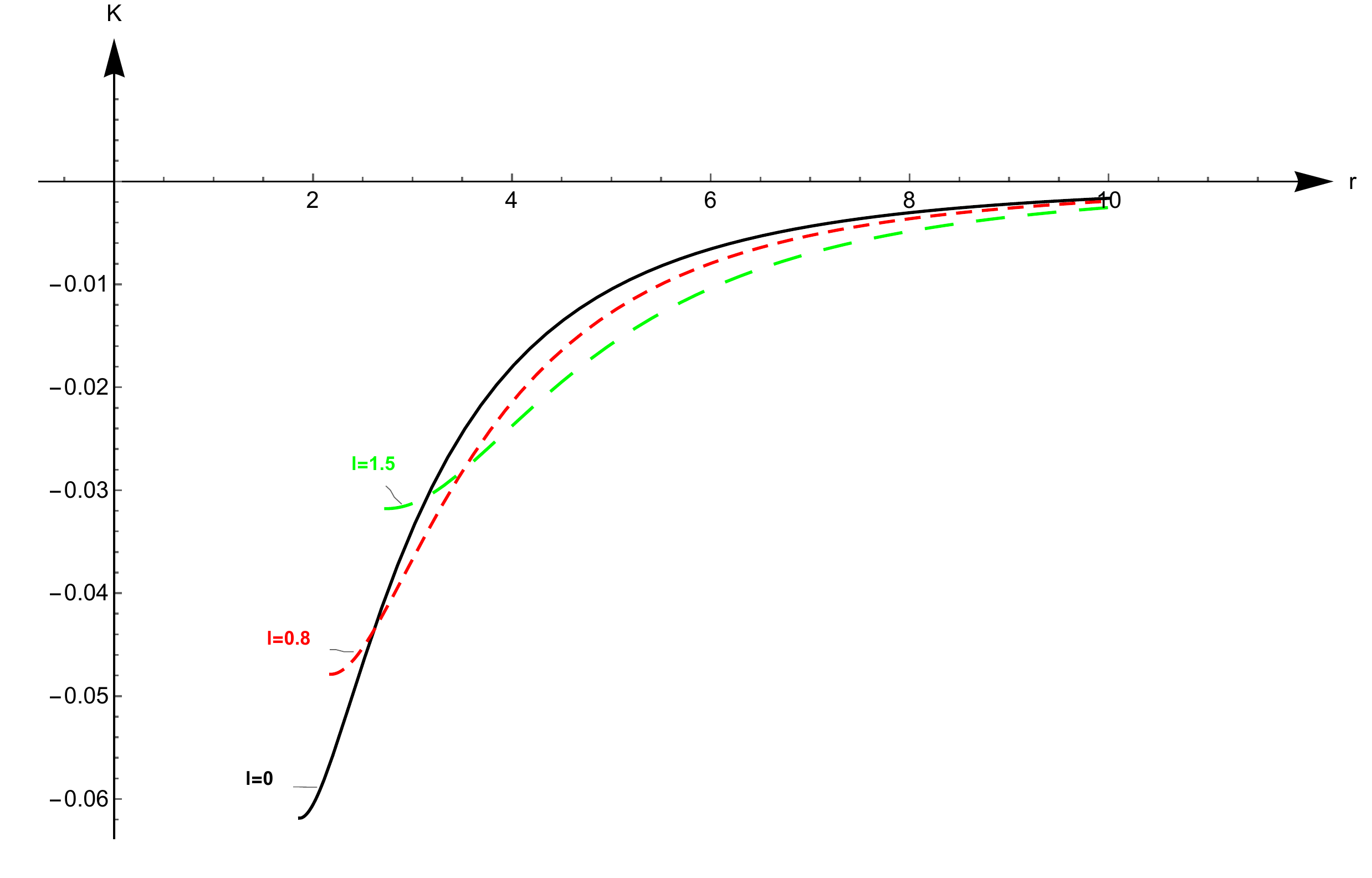}
    \caption{The Gaussian curvature of the optical metric for different NUT parameters,
    plotted in each case for $m + \sqrt{m^2 + l^2-e^2} < r$ and $e=0.5$, in units with $m=1$.}
    \label{cBlack}
\end{figure}

As a matter of fact, from Eq. (\ref{eq:K}) we can deduce that the Gaussian curvature $K(r)$ of the optical metric is negative in the entire domain $m + \sqrt{-b} < r < \infty$, not only near $r=\infty$ and not only on the embeddable part. To prove this, one may substitute $r = m + \sqrt{-b}+\xi$. Then $(r^2+\ell^2)^4 K(r)$ becomes a fifth-order polynomial in $\xi$. All coefficients of this polynomial are manifestly negative, so $K(r)$ is indeed negative for $\xi >0$, i.e., on the domain of outer communication. The negative sign of $K(r)$ implies that the geodesics of the optical metric locally diverge. Fig. \ref{cBlack} shows plots of the Gaussian curvature as a function of $r$.

The fact that the Gaussian curvature is negative has an important consequence if we take the Gauss-Bonnet theorem into account. It was shown in a pioneering paper by Gibbons and Werner\cite{Gibbons:2008rj} that in spherically symmetric and static spacetimes that are asymptotically flat the deflection angle can be written as an area integral over the negative of the Gaussian curvature of the optical metric which, in this case, lives on a (coordinate) plane. Halla and Perlick \cite{M1} have demonstrated that the same is true in the NUT metric, where now the optical metric lives on a (coordinate) cone. The construction carries over, without any modification, to the case of Brill black holes ($b \le 0$) with $e \neq 0$. So also in this case the deflection angle is an area integral over the negative of the Gaussian curvature of the optical metric. Therefore, our result that the Gaussian curvature is negative implies that the deflection angle is positive, i.e., that a light ray is deflected towards the center, as shown in Fig.~\ref{rm}. This is not obvious from the line-integral formula for the deflection angle.


\section{Gravitational lensing of Brill wormholes} \label{NUTW}   
In this section we consider a Brill spacetime with $m \ge0$, $l\neq 0$ and $b>0$. As there are no horizons and no singularities, the radial coordinate ranges over all of $\mathbb{R}$. In the first part, we discuss the existence of photon spheres and their relevance for the lensing features. In the second part, we check if the cones $\vartheta = \mathrm{constant}$ with the optical metric (\ref{eq:opt}) can be isometrically embedded in Euclidean 3-space for the entire domain $-\infty<r<+\infty$ and whether the Gaussian curvature of the optical metric is always negative.

\subsection{Photon spheres, shadow and deflection angle}
\begin{figure}[H]
  \centering
  {\includegraphics[width=0.45\linewidth]{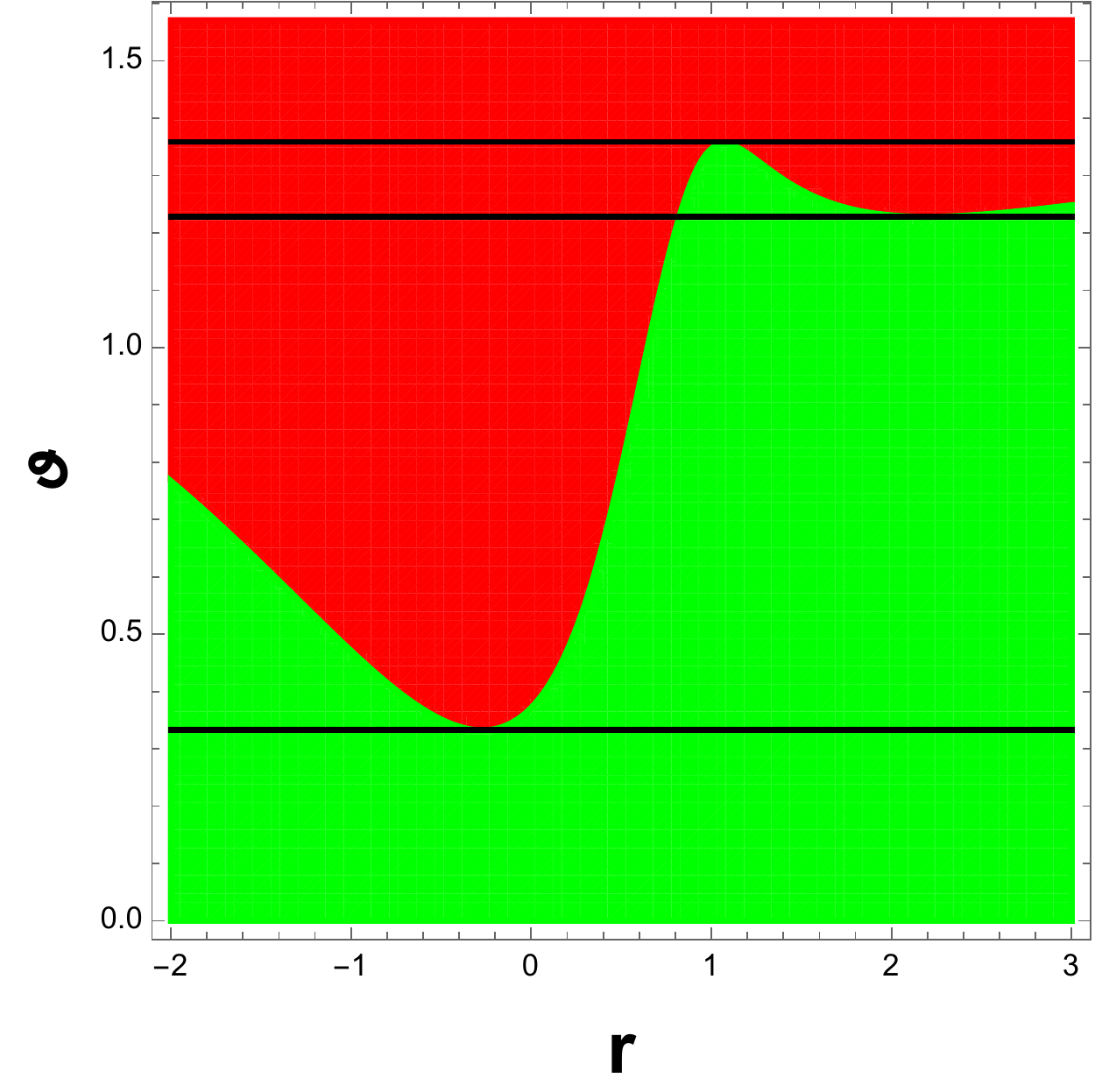}}%
  \caption{The allowed region $V_{\vartheta}(r) \le - 4 l^2 b$ (green) for light rays in Case 1), with $l=0.8 m$ and $e=1.3 m$. $r$ is given in units of $m$.}
  \label{P1}
\end{figure}

\begin{figure}[H]
  \centering
  {\includegraphics[width=0.45\linewidth]{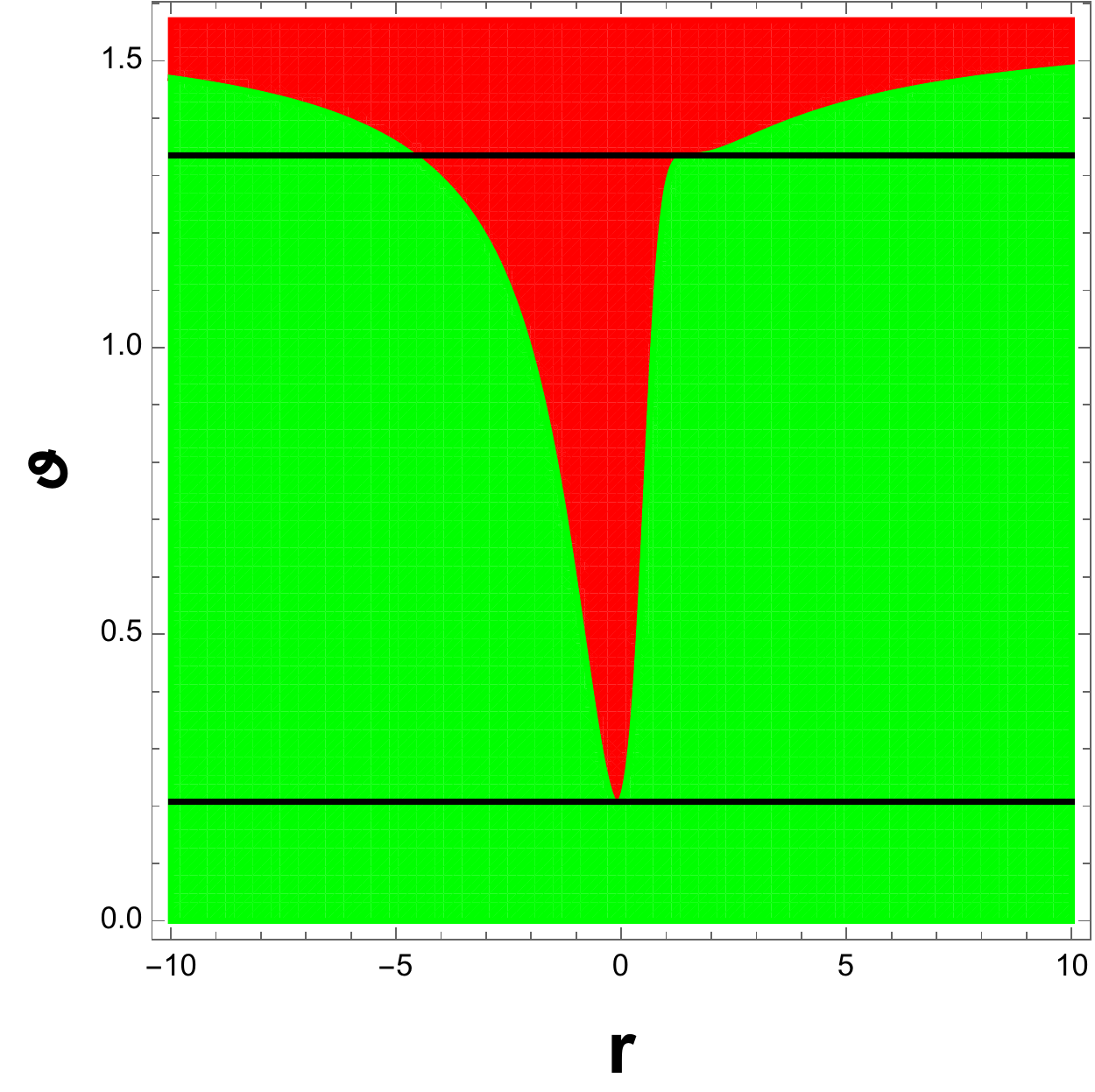}}%
  \caption{The allowed region $V_{\vartheta}(r) \le - 4 l^2 b$ (green) for light rays in Case 2), with $l=0.46m$ and $e=1.17 m$.  $r$ is given in units of $m$.}
  \label{P2}
\end{figure}

\begin{figure}[H]
  \centering
  {\includegraphics[width=0.45\linewidth]{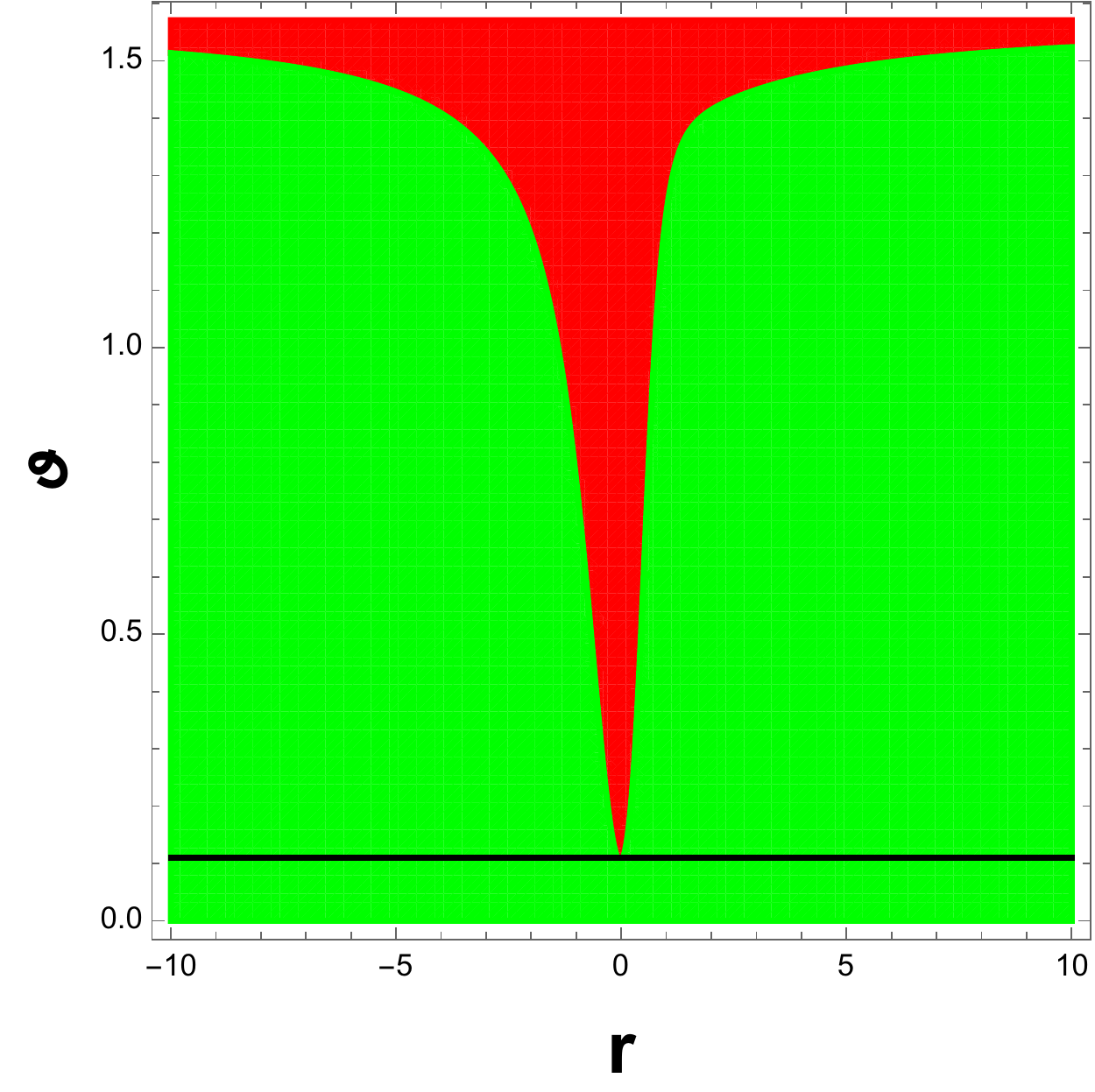}}%
  \caption{The allowed region $V_{\vartheta}(r) \le - 4 l^2 b$ (green) for light rays in Case 3), with $l=0.26 m$ and $e=1.26 m$. $r$ is given in units of $m$. }
  \label{P3}
\end{figure}

\begin{figure}[H]
  \centering
  {\includegraphics[width=0.45\linewidth]{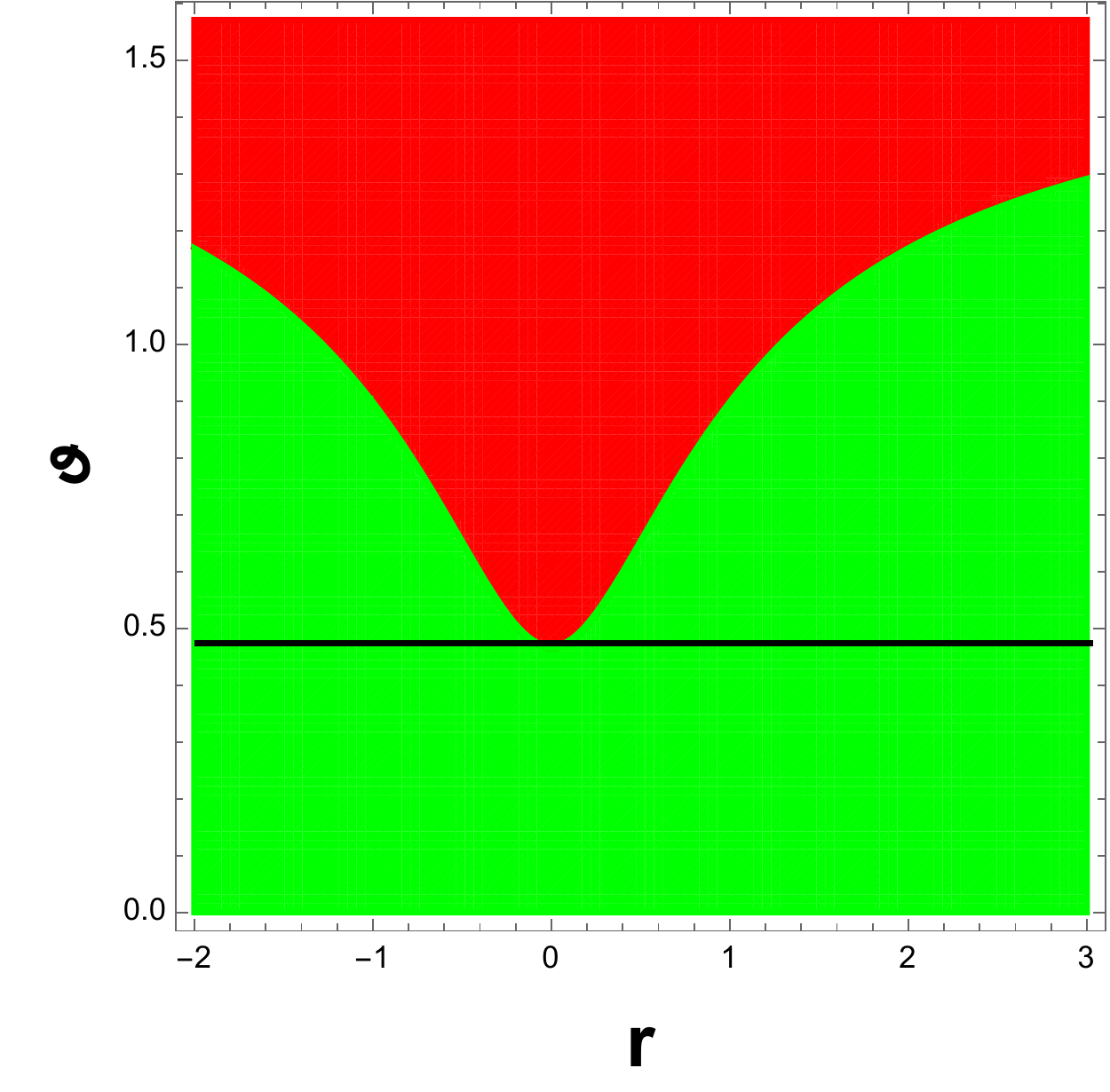}}%
  \caption{The allowed region $V_{\vartheta}(r) \le - 4 l^2 b$ (green) for light rays in Case 4), where $m=0$, $l=0.435 r_0$ and $e=0.611 r_0$. $r_0$ is a constant that has the dimension of a length and $r$ is given in units of $r_0$. }
  \label{P4}
\end{figure}

Recall that the radius coordinate of a photon sphere is determined by the cubic equation (\ref{eq:rph}). In the wormhole case $b>0$ considered here the
number of real solutions depends on the discriminant 
\begin{equation}
    \Delta = (m^2+l^2-2b)^3-27 m^2 b^2 \, .
\end{equation}
\begin{figure}[H]
\centering
    \includegraphics[width=0.7\textwidth]{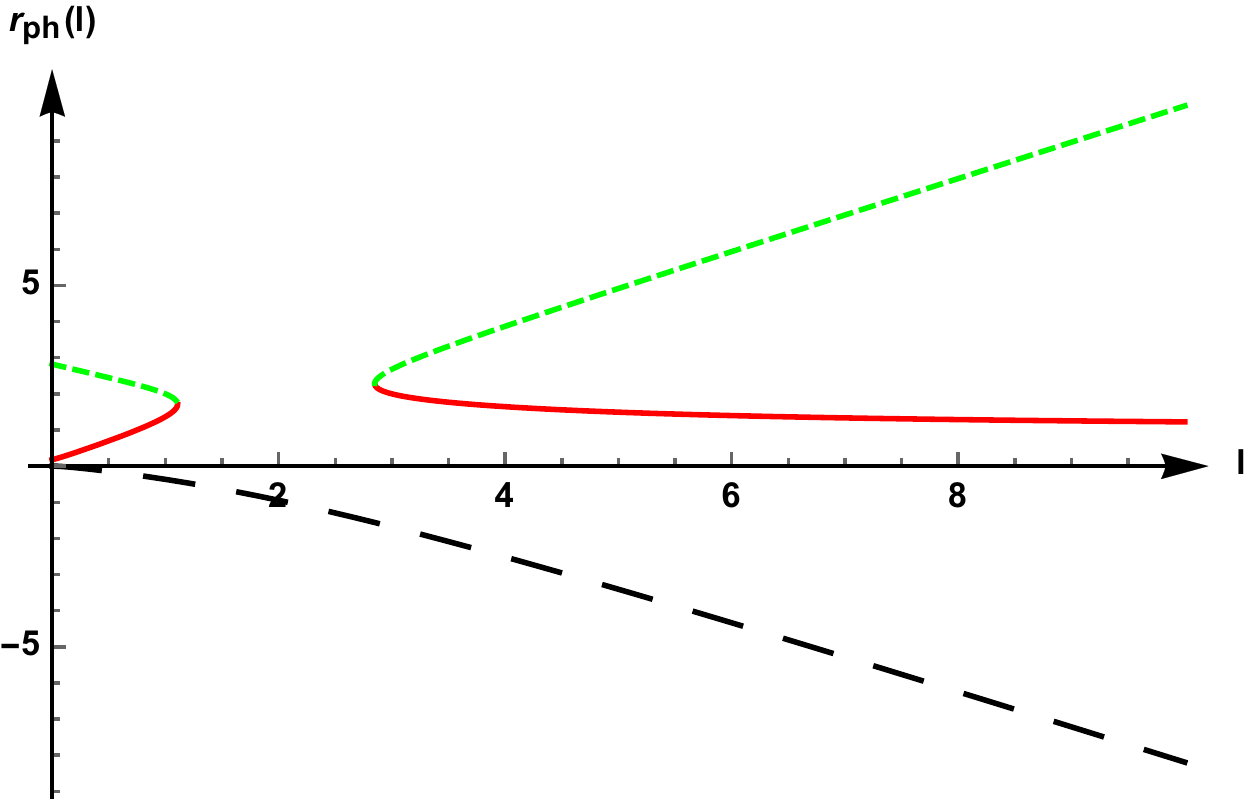}
    \caption{Solutions of Eq. \eqref{eq:rph} with $e= (2l+m)/2$, plotted in units of $m$.}
    \label{c}
\end{figure}
We distinguish four cases.
\begin{itemize}
\item \textbf{Case 1) $\boldsymbol{\Delta > 0}$}. Then there are three distinct real roots. With the help of the effective potential (\ref{potV}) one sees that two of them are unstable and the one between them is stable. If we label the three roots such that $r_{\mathrm{ph}1}< r_{\mathrm{ph}2}<r_{\mathrm{ph}3}$, one can deduce from (\ref{eq:rph}) and (\ref{eq:thetaph}) that the corresponding opening angles satisfy $\vartheta_{\mathrm{ph}1}<\vartheta_{\mathrm{ph}3}<\vartheta_{\mathrm{ph}2}$.   In Fig.~\ref{c}, $r_{\mathrm{ph1}}$ corresponds to the black (wide-dashed) curve, $r_{\mathrm{ph2}}$ corresponds to the red curve and $r_{\mathrm{ph3}}$ corresponds to the green (dashed) curve.
\item \textbf{Case 2) $\boldsymbol{\Delta =0}$ and $\boldsymbol{m >0}$}. Then there are three real roots two of which coincide. The photon sphere at the single root is unstable whereas the one at the double root is marginally stable. This case is the limit of Case 1) where $r_{\mathrm{ph}2}=r_{\mathrm{ph}3}$. The marginally stable photon sphere is where the red curve and the green (dashed) curve come together in Fig.~\ref{c}.
\item \textbf{Case 3) $\boldsymbol{\Delta < 0}$}. Then there is one real root. The other two roots are non-real and complex conjugate to each other. In this case there is only one photon sphere and it is unstable. Its radius value is given by the black (wide-dashed) curve in Fig.~\ref{c}.
\item \textbf{Case 4) $\boldsymbol{\Delta = 0}$ and $\boldsymbol{m = 0}$}. Then all three roots are real and coincide. There is one photon sphere and it is unstable. For this case a plot analogous to Fig.~\ref{c} would show only the black (wide-dashed) curve.  
\end{itemize}
Recall that lightlike geodesics exist in the region where the effective potential (\ref{potV}) satisfies $V_{\vartheta }(r) \le -4 l ^2 b$. In analogy to Fig.~\ref{P1_black}, this region in the $r-\vartheta-$plane is marked in green in Figs. \ref{P1}, \ref{P2}, \ref{P3} and \ref{P4}, which correspond to Cases 1), 2), 3) and 4), respectively. Each lightlike geodesic runs on a horizontal straight line in the green region. The forbidden region  is marked in red. With the help of these figures, one can nicely discuss the different types of lightlike geodesics.

Fig. \ref{P1} corresponds to Case 1). The boundary curve between the green and the red regions has two minima and one maximum. The minima correspond to unstable photon spheres, the maximum corresponds to a stable photon sphere. The horizontal black lines mark, from bottom to top,  the opening angles $\vartheta _{\mathrm{ph}1}$, $\vartheta _{\mathrm{ph}3}$,  and $\vartheta _{\mathrm{ph}2}$.  From this figure we read that, in addition to light rays completely contained in a photon sphere, the following types of light rays exist: 
\begin{itemize}
\item \textbf{Case (a)} There are light  rays that run from  $+ \infty$ to $-\infty$ or vice versa. They lie in cones with opening angles $\vartheta < \vartheta _{\mathrm{ph}1}$.
\item \textbf{Case (b)} There are light rays that come from $+ \infty$ or $- \infty$, are reflected at the boundary of the green area and escape back to  $+ \infty$ or $- \infty$.  They lie in cones with opening angles $\vartheta _{\mathrm{ph}1}< \vartheta$ and $\vartheta \neq \vartheta _{\mathrm{ph}3}$.
\item \textbf{Case (c)} There are light rays that oscillate back and forth between a maximum and a minimum value of their radius coordinate. They lie in cones with opening angles $\vartheta _{\mathrm{ph}3}<\vartheta < \vartheta _{\mathrm{ph}2}$. These light rays oscillate around the stable photon sphere at $r_{\mathrm{ph2}}$.
\item \textbf{Case (d)} The limiting cases between (a) and (b) or between (b) and (c) are light rays which asymptotically spiral towards a photon circle in one of the two unstable photon spheres.
They lie in cones with opening angles $\vartheta = \vartheta _{\mathrm{ph}1}$ or $\vartheta = \vartheta _{\mathrm{ph}3}$, i.e., they run along the lowest or the middle black line in Fig.~\ref{P1}. Therefore, there are four different situations possible: 
\begin{itemize}
\item A light ray comes from $- \infty$ and asymptotically approaches an unstable photon circle at $r _{\mathrm{ph}1}$, or vice versa.
\item A light ray comes from $+ \infty$ and asymptotically approaches an unstable photon circle at $r _{\mathrm{ph}1}$, or vice versa.
\item A light ray comes from $+ \infty$ and asymptotically approaches an unstable photon circle  at $r _{\mathrm{ph}3}$, or vice versa.
\item A light ray starts asymptotically at an unstable photon circle at $r_{\mathrm{ph}3}$, goes through a minimal $r$ value and then asymptotically approaches the same photon circle from which it started. Such orbits are called \emph{homoclinic}. 
\end{itemize}
\end{itemize}
Analogously, the behaviour of light rays in Cases 2), 3) and 4) can be read from Figs. \ref{P2}, \ref{P3} and \ref{P4}. In Case 2) the unstable photon sphere at $\vartheta _{\mathrm{ph}3}$ merges with the stable photon sphere at $\vartheta _{\mathrm{ph}2}$ which results in a marginally stable photon sphere. Correspondingly, Case (c) and the last subcase of Case (d) are no longer possible. 
In Case 3) and in Case 4) the photon spheres at $\vartheta _{\mathrm{ph}3}$ and $\vartheta _{\mathrm{ph}2}$ are gone, so only Case (a), Case (b) and the first two subcases of Case (d) are possible.

\begin{figure}[H]
\centering
    \includegraphics[width=0.8\textwidth]{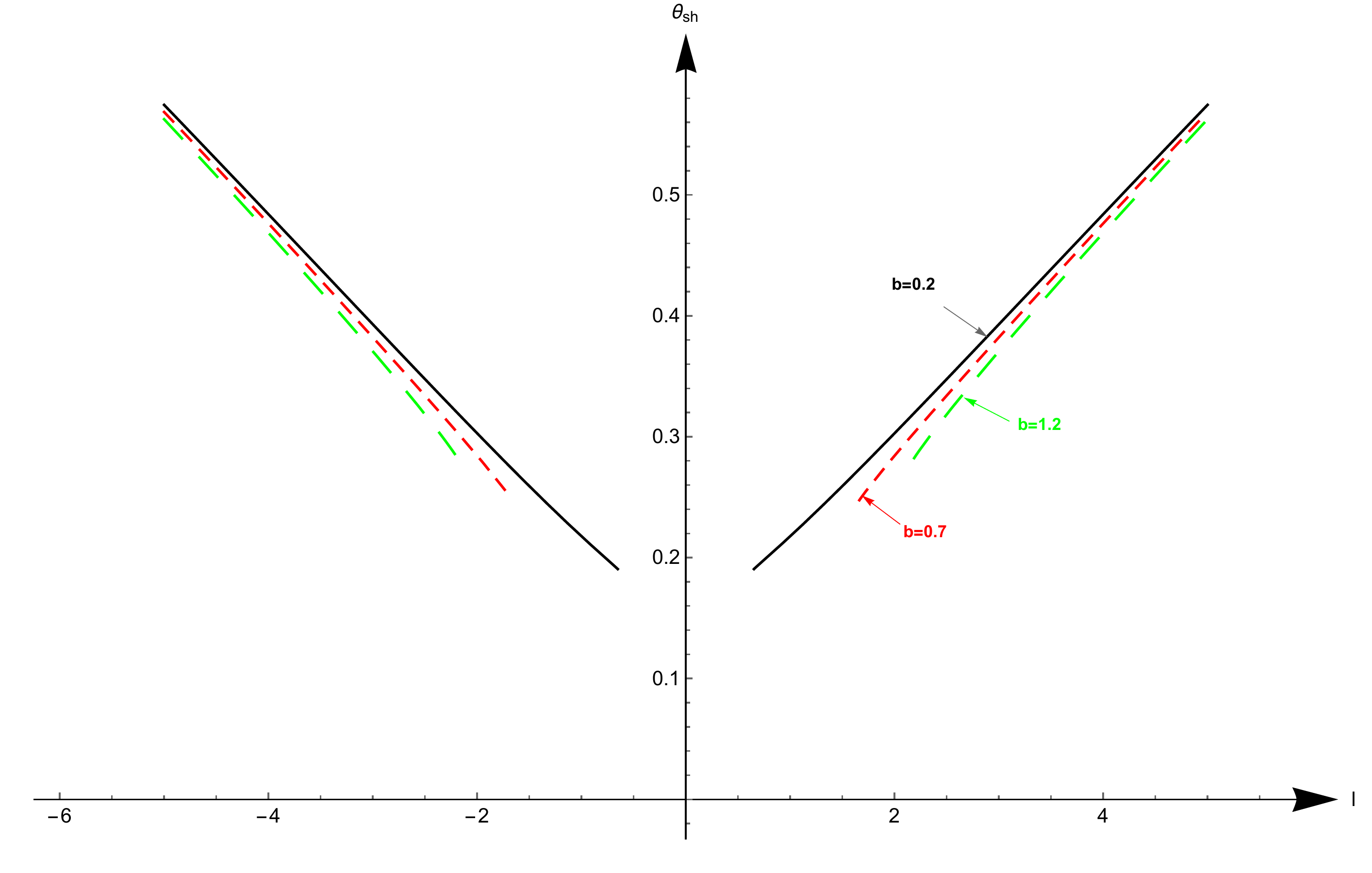}
    \caption{ Shadow radius of a Brill wormhole, for an observer at $r_{\mathrm{O}}=20m$, with $l$ and $b$ given in units with $m=1$.} 
    \label{fig:sh2}
\end{figure}

Brill wormholes cast a shadow, similarly to Brill black holes. Its angular radius $\theta _{\mathrm{sh}}$ is again given by Eq. (\ref{eq:shadowbh}) where now $r_{\mathrm{ph1}}$ has to be read for $r_{\mathrm{ph}}$ in Case 1) and Case 2). $r_{\mathrm{O}}$ is the radius coordinate of the observer who is assumed to be stationary anywhere in the wormhole spacetime. The shadow is again a circular black disk in the sky if we assume that there are light sources anywhere but not in the region bounded by the past-oriented light rays that spiral from the observer position to the photon sphere at $r_{\mathrm{ph1}}$ in Case 1) and Case 2) or to the unique photon sphere in Case 3) and Case 4). In the wormhole case considered                  here this region extends to $r= - \infty$. Note that here our assumption on the position of light sources is essential. As an alternative, we could assume that we have light sources only at big positive radius values. This would make a difference in those cases where several photon spheres exist. In Case 1), for an observer position in the region where light rays exist that oscillate around the stable photon sphere, we associate brightness with these light rays. However, one would associate darkness with them if light sources are only at big positive radius values. So in the latter case the shadow would consist not only of a dark disk but in addition of a dark ring around this disk. For an observer in Case 1) at the boundary of the region where oscillating light rays exist, and for an observer at the marginally stable photon sphere in Case 2), this (two-dimensional) ring in the sky would reduce to a (one-dimensional) circle. Fig~\ref{fig:sh2} shows the dependence on the parameters $l$ and $b$ of the angular radius of the shadow of a wormhole, for an observer at fixed radius coordinate.

The deflection angle is well-defined only for light rays of Case (b). All formulas for the deflection angle can be literally taken over from the black-hole case, see Section \ref{black}, both for light rays that come in from $r=+ \infty$ and go back to $r=+\infty$ and for light rays that come in from $r=-\infty$ and go back to $r=-\infty$. We just have to keep in mind that now $r$ may take negative values and that for a light ray coming from $r=- \infty$ and going back to $r=-\infty$ there is no minimum radius $r_m$ but rather a maximum radius $r_m$. With these adjustments, all formulas from Section \ref{black} remain valid. In particular, the deflection angle is still given by an integral over the negative of the Gaussian curvature of the optical metric. We will see in the next subsection that in the wormhole case this Gaussian curvature may be positive, so the deflection angle may be negative which means that the light ray is repelled from the center. In such a situation the angle $\Delta \tilde{\varphi}$ in Fig.~\ref{rm} is smaller than $\pi/2$. We will also discuss the related embedding diagrams in the next subsection.

\subsection{Embedding diagrams}
\begin{figure}[H]
  \centering
  \subfloat[][]{\includegraphics[width=0.5\linewidth]{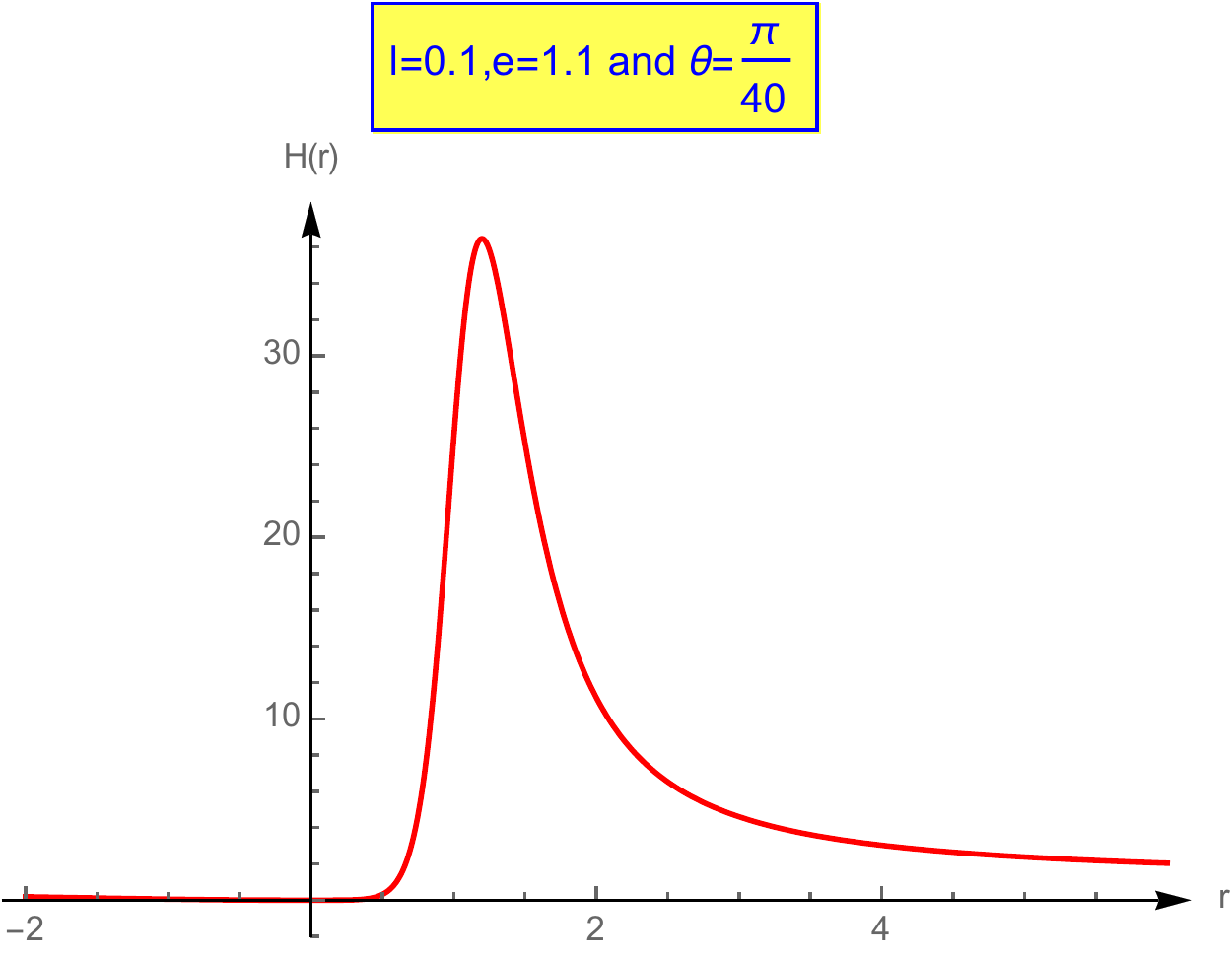}}%
  \qquad
  \subfloat[][]
  {\includegraphics[width=0.5\linewidth]{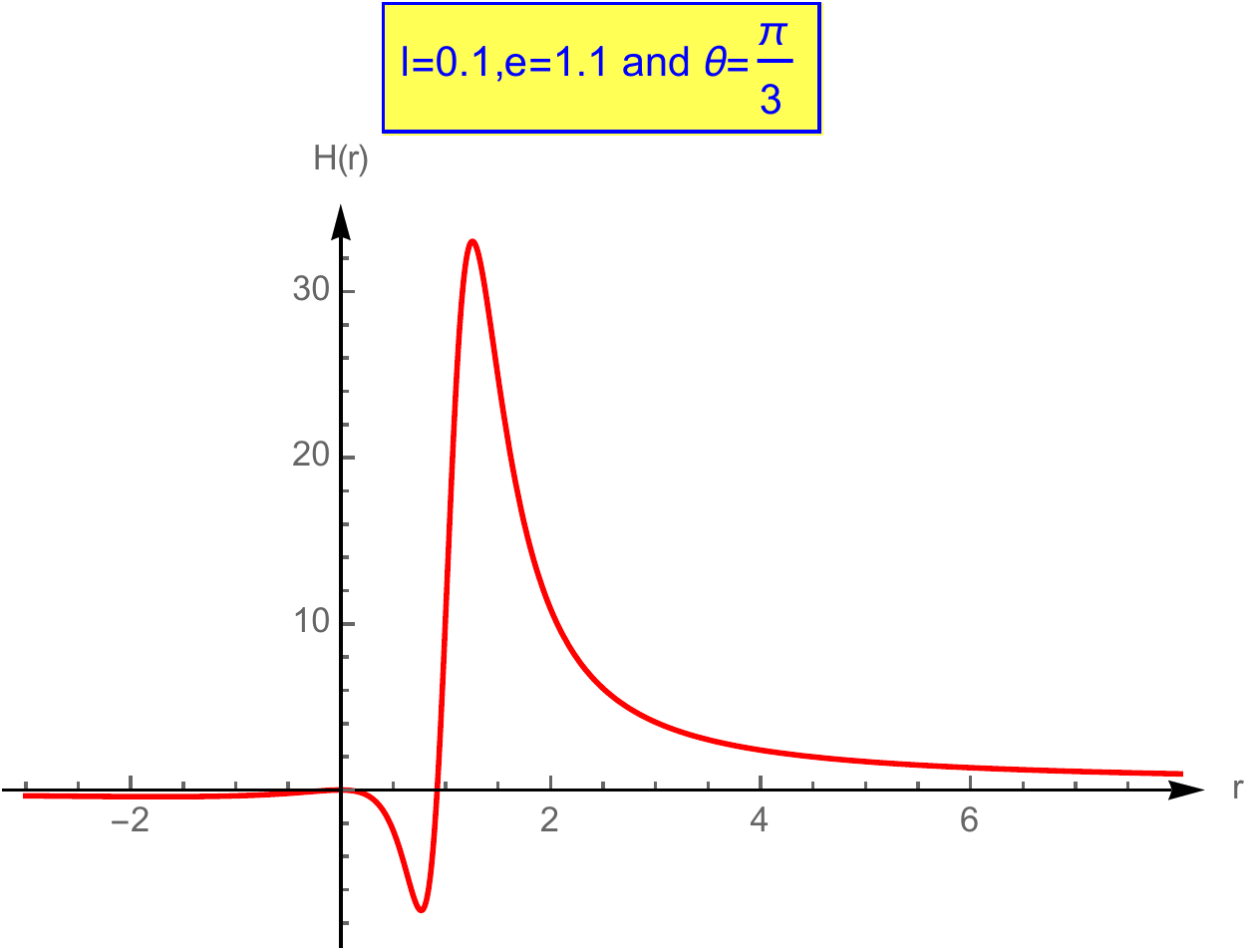}}%
  \caption{In (a), H is positive, and in (b), the positivity of $H$ is violated for larger $\vartheta$. The plot is in units with $m=1$.}
  \label{H}
\end{figure}

\begin{figure}[H]
\centering
  {\includegraphics[width=0.3\linewidth]{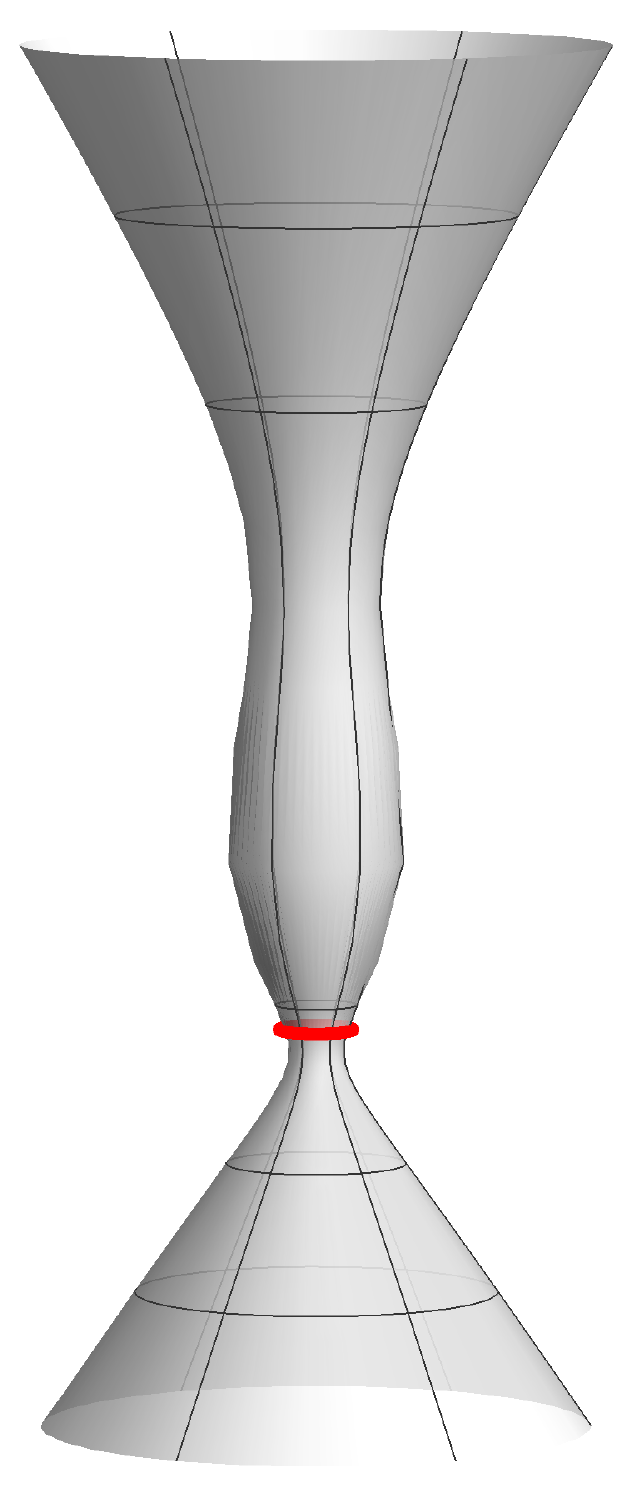}}%
  \caption{The cone $\vartheta = \vartheta _{\mathrm{ph1}}=0.563$ of the optical metric (\ref{eq:opt}) with $l=2.3m$ and $e=2.6m$, embedded into Euclidean 3-space as a surface of revolution and with a photon circle (red) at $r=r_{\mathrm{ph1}}=-1.396m$.}
  \label{embworm1}
\end{figure}

\begin{figure}[H]
\centering
  {\includegraphics[width=0.3\linewidth]{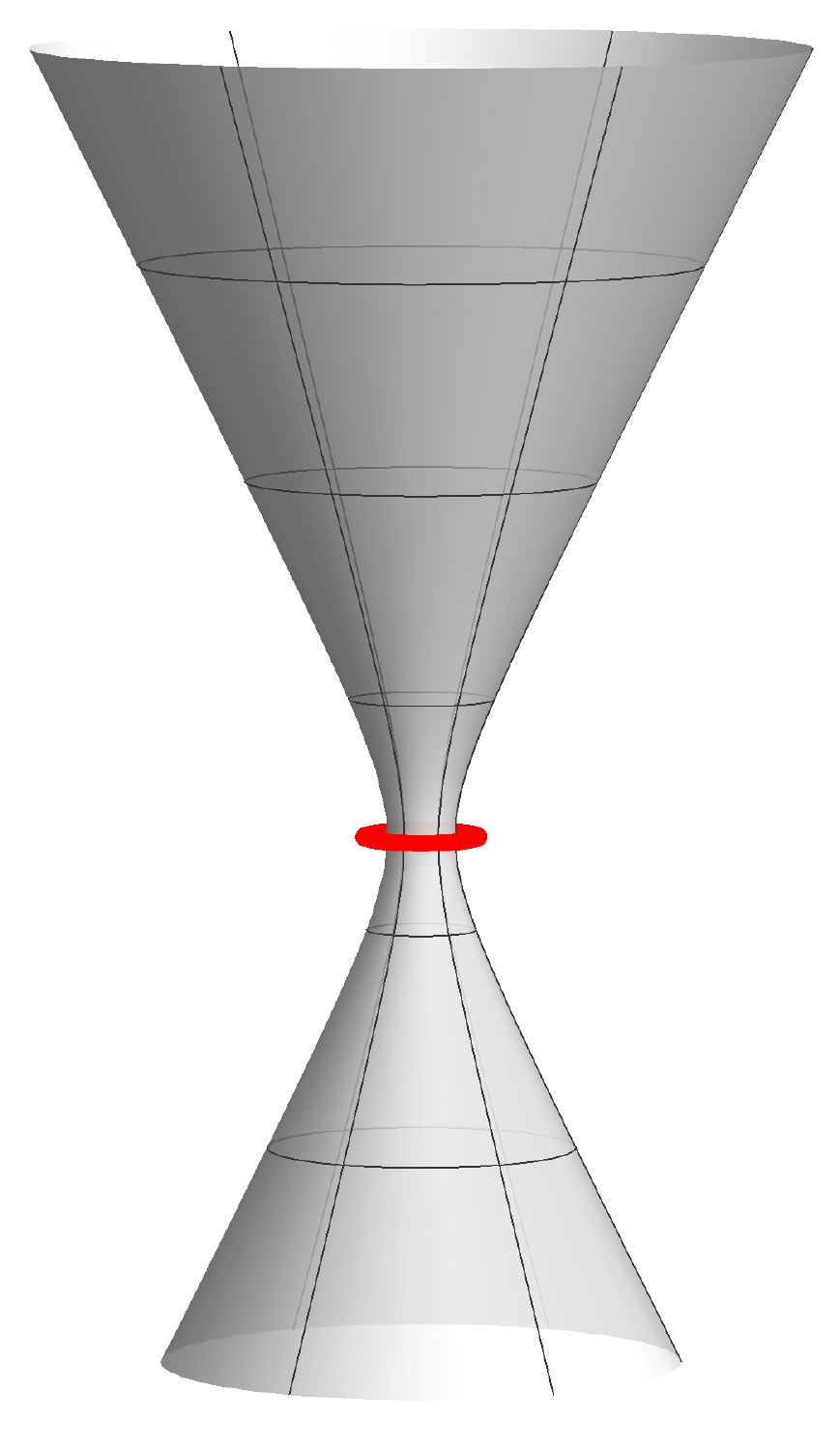}}%
  \caption{The cone $\vartheta = \vartheta _{\mathrm{ph}}=0.467$ of the optical metric (\ref{eq:opt}) with $m=0$, $l=1.3 r_0$ and $e= 1.83 r_0$, embedded into Euclidean 3-space as a surface of revolution and with a photon circle (red) at $r=r_{\mathrm{ph}}=0$. $r_0$ is a positive constant with the dimension of a length.}
  \label{massless}
\end{figure}

In analogy to what we have done for the black-hole case in Sec. \ref{black}, we will now isometrically embed the coordinate cone $\vartheta=$ constant into Euclidean 3-space for the wormhole case. From Eq. (\ref{eq:emb}) we read that this is possible provided that the function $H(r)$ is non-negative. In Fig. \ref{H}, we plot $H(r)$ for different values of $l$, $e$ and $\vartheta$. This exemplifies that the condition of embeddability is not in general satisfied on the entire domain $-\infty < r < \infty$. 
In Figs. \ref{embworm1} and \ref{massless} we show two examples where the embeddability condition is satisfied on this entire domain. In both pictures we show a cone $\vartheta=$ constant that contains an unstable photon circle. As is exemplified in Fig.~\ref{embworm1}, in contrast to black holes (see Fig. \ref{em1}) for Brill wormholes not only local minima (``necks'') but also local maxima (``bellies”) are possible. In the neighborhood of a local maximum the Gaussian curvature is positive which implies that on this neighborhood light rays locally converge. This is in agreement with our earlier observation that there may be stable photon spheres in a Brill wormhole spacetime. As a matter of fact, the wormhole spacetime of Fig.~\ref{embworm1} is of Case 1), so there is a second unstable photon sphere and also a stable photon sphere. However, their corresponding opening angles are different from the one to which the shown embedding diagram applies. In Fig.~\ref{embworm1} the Gaussian curvature is positive not only near the ``belly'' but also near $r=-\infty$ which implies that the deflection angle is negative for light rays that come in from $r=-\infty$ and go back to $r=-\infty$, i.e., that such light rays are repelled from the center. -- By contrast, in Fig.~\ref{massless} we have chosen an example where the Gaussian curvature is everywhere negative. As in this example $m=0$, the spacetime geometry is symmetric with respect to reflections $r \mapsto -r$. There is only one photon sphere, which is situated at the ``neck'' at $r=0$ and unstable. So this example is more similar to the usual wormhole geometry than the one shown in Fig.~\ref{embworm1}.


We have already mentioned that the Gaussian curvature $K(r)$, given by Eq. (\ref{eq:K}), is not necessarily negative everywhere in the wormhole case. We know already from Eqs. (\ref{eq:TaylorK1}) to (\ref{eq:TaylorRZ3}) that the geometry approaches that of a (flat) cone for $r \to \pm \infty$ and that the asymptotic cone is approached from the outside if $K(r)$ is negative and from the inside if $K(r)$ is positive near $\pm \infty$. If $m \neq 0$, by Eq. (\ref{eq:TaylorK1}) $K(r)$ is negative near $r = + \infty$ and positive near $r=- \infty$. If $m=0$ and $3e^2 \neq 7 l^2$ we read from Eq. (\ref{eq:TaylorK2}) that $K(r)$ is positive near $r=+\infty$ and $r=-\infty$ if $3e^2>7 l^2$ and negative near $r=+\infty$ and $r=-\infty$ if $3e^2<7l^2$. Both cases are compatible with the wormhole condition $b>0$. Finally, if $m=0$ and $3e^2=7l^2$, have to use Eq. (\ref{eq:TaylorK3}) which tells us that $K(r)$ is negative near $r=+\infty$ and also near $r=- \infty$. Recall that, by the Gauss-Bonnet theorem, $K(r)$ being positive on a region near $r = \pm \infty$ means that the deflection angle $\delta$ is negative for light rays that stay within this region, i.e., that such a light ray is repelled from the center.  

\begin{figure}[H]
  \centering
  \subfloat[][]
  {\includegraphics[width=0.3\linewidth]{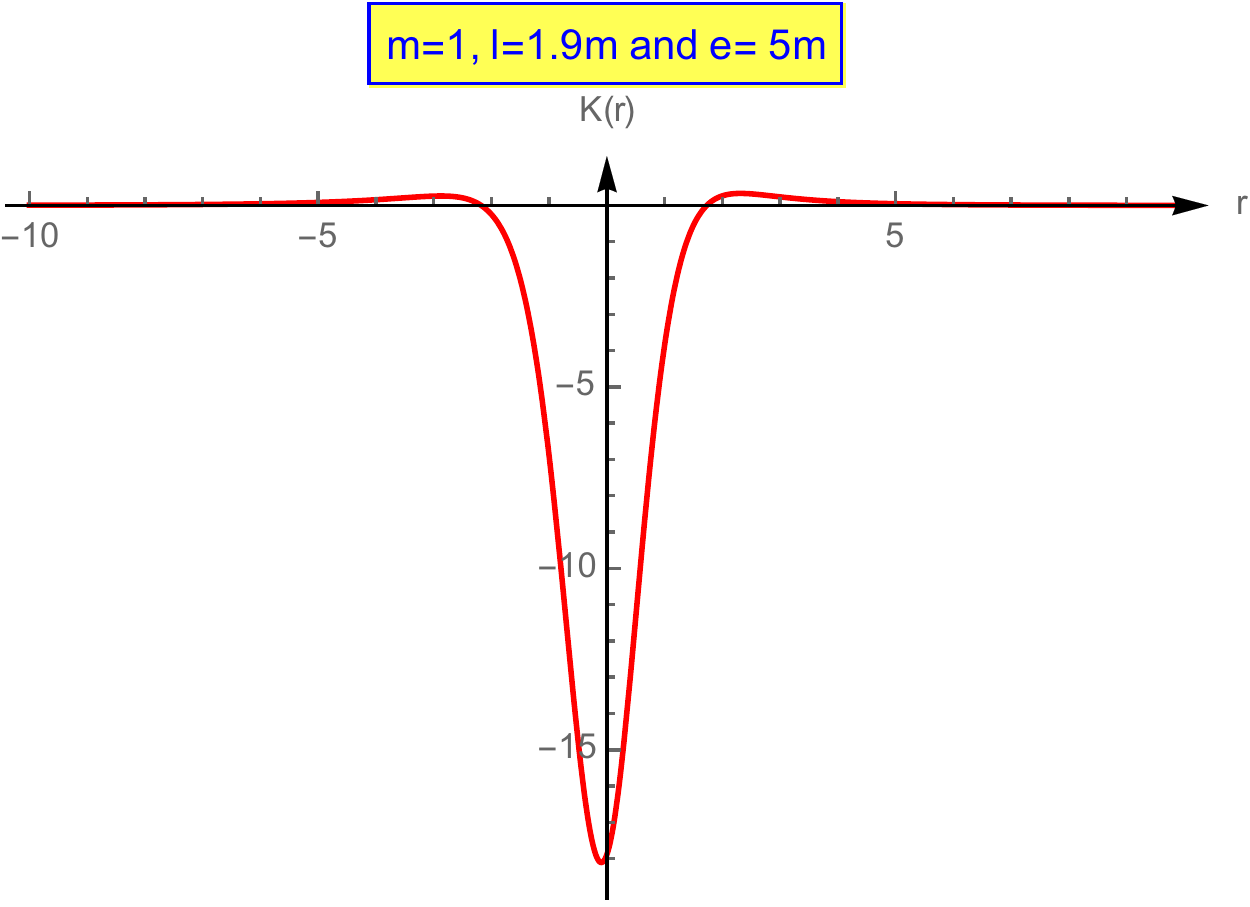}}%
  {\includegraphics[width=0.3\linewidth]{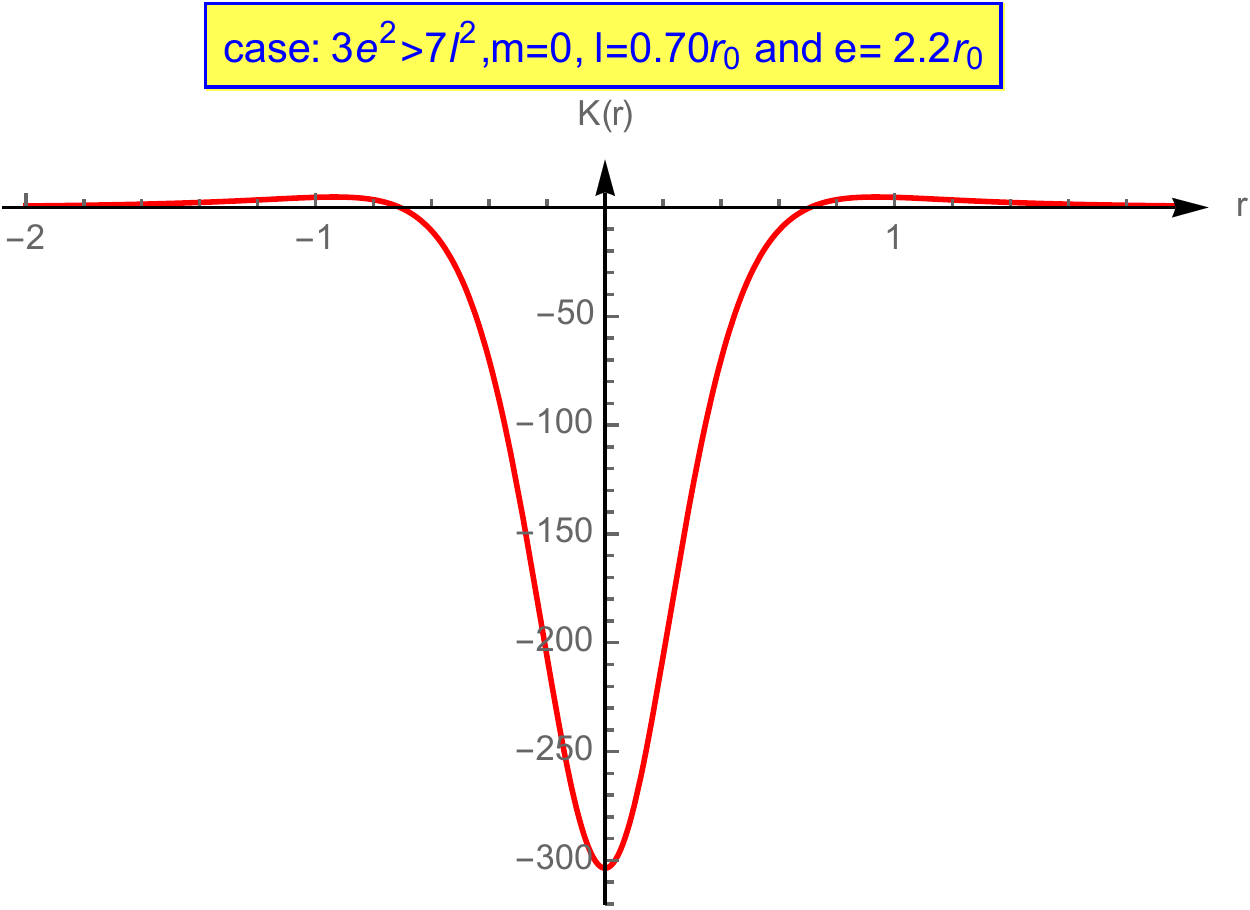}}%
  {\includegraphics[width=0.3\linewidth]{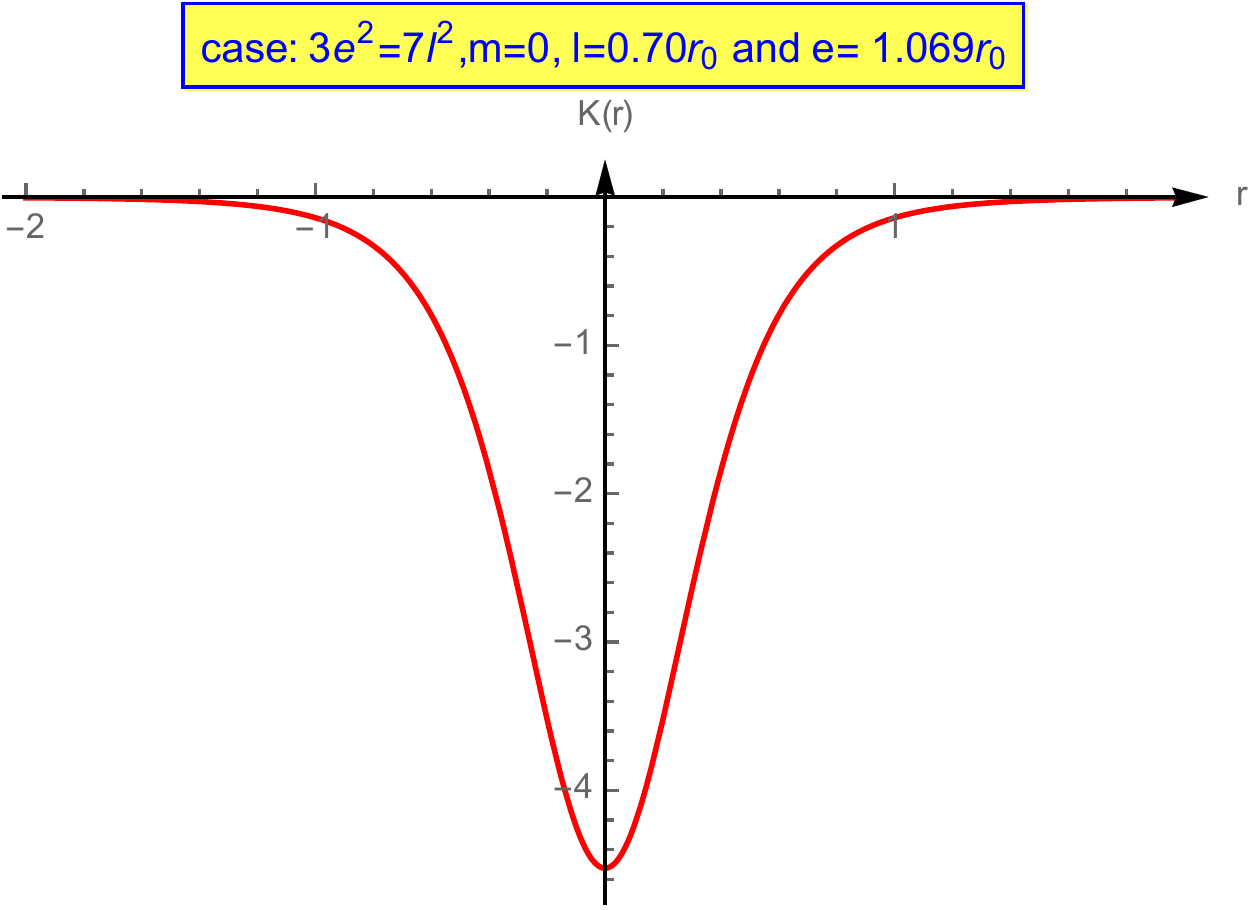}}%
  \caption{Plot of the Gaussian curvature $K$ from Eq. (\ref{eq:K}) as function of $r$ on the entire domain $-\infty<r<+\infty$ with different values of $m$, $e$ and $l$.}
  \label{embworm}
\end{figure}

In Fig. \ref{embworm} we plot the Gaussian curvature represented by Eq. (\ref{eq:K}). This figure exemplifies the observation that $K(r)$ can be positive within certain intervals of $r$. This can also be noticed in the embedding diagram.   
\section{CONCLUDING REMARKS}\label{sec:conclusion}
In this paper we have discussed the lensing features of Brill spacetimes. In particular, we have worked out the relevant formulas for the photon spheres, for the shadow and for the deflection angle. We have also shown that every light ray is contained in a (coordinate) cone and that, on this cone, it is a geodesic of a Riemannian optical metric. This allows to determine the sign of the deflection angle with the help of the Gauss-Bonnet theorem and to visualize several lensing features in terms of embedding diagrams. Thereby one has to distinguish two types of Brill spacetimes that are physically quite different: If the parameter we called $b$ is smaller than or equal to 0, the spacetime describes a black hole; however, if this parameter is bigger than 0 it describes a wormhole. We have seen that in the black-hole case the lensing features are qualitatively quite similar to those of a NUT black hole which had been treated earlier by Zimmerman and Shahir \cite{ZimmermanShahir1989} and by Halla and Perlick \cite{M1}, i.e., the charge parameter changes these features only quantitatively. The wormhole case, however, is more complicated, in particular because there may be several photon spheres and the Gaussian curvature of the optical metric need not be negative everywhere. It is true that there is no observational evidence for the existence of Brill wormholes (or Brill black holes with non-zero NUT or charge parameter) in Nature. However, their existence cannot be ruled out and several people find them interesting because within Einstein's general relativity theory they are the only known traversable wormholes without exotic matter. For this reason we believe that it was worthwhile to derive their lensing features in detail.                 

\begin{acknowledgments}
We gratefully acknowledge support from the DFG within the Research Training Group 1620 “Models of Gravity”.
\end{acknowledgments}
\bibliographystyle{unsrtnat}

\end{document}